\documentclass[]{interact}

\usepackage[caption=false]{subfig}

\usepackage[natbibapa,nodoi]{apacite}
\theoremstyle{plain}

\theoremstyle{definition}

\theoremstyle{remark}

\begin{document}

\articletype{ARTICLE}

\title{\textit{NoteG}: A Computational Notebook to Facilitate Rapid Game Prototyping}

\author{
\name{Noble Saji Mathews\thanks{CONTACT Noble Saji Mathews Email: ch19b023@iittp.ac.in} and Sridhar Chimalakonda}
\affil{Research in Intelligent Software \& Human Analytics Lab, Indian Institute of Technology Tirupati, India}
}


\maketitle

\begin{abstract}
Game development-based approaches are increasingly used to design curricula that can engage students, as these can help them apply and practice learnt computer science concepts. 
However, it can become complex to develop a minimum working game or a prototype with the help of high-end game engines. Game prototyping is one of the most essential parts of the game design and development cycle as it allows developers to continuously test and improve their ideas. 
In recent years, computational notebooks have gained widespread popularity among developers. They can help run individual code snippets, visualize the output, consolidate the source code, and share live code easily. However, its use has not been explored in the field of game development and prototyping. In this paper, we propose \textit{NoteG}, a computational notebook towards rapid game prototyping. We evaluated the tool with 18 novice game developers through a questionnaire-based user survey. A majority of the volunteers (66\%) found it easy to use and were of the opinion that it saves time. 
A few of the participants successfully extended the existing framework to implement new game mechanics within their prototypes. 
\end{abstract}


\section{Introduction}

Gamification is being increasingly used in various domains ranging from skill identification in recruitment to building engaging course curricula in fields such as computer science, mathematics and biology
\citep{obaid2020gamification, 10.1145/3304221.3319792,schulz2020user,lo2020comparison,siswati2021developing}. Students often show a keen interest in gamified approaches, and hence they are being explored to enhance traditional programming education \citep{10.1145/2729094.2742590}.
There have also been explorations into encouraging children to develop games on their own \citep{10.1145/2771839.2771926}. Game development based approaches have been found to be effective in promoting problem-solving skills among children \citep{chu2015effects}. With more and more students being exposed to game development or even taking it up based on their own interests \citep{rahimi2019role}, there is a need for the inclusion of proper game design practices in these initial stages and tools to simplify the process. 

Challenges in building a game that can implement the desired functionality are often attributed to the tools used and workflow involved  \citep{10.1145/3365000}. There have been efforts that address this issue by shifting the focus to mobile-based game development, which is considered to be less complex so that students with limited experience are also able to create playable games \citep{10.1145/1539024.1508881}. 

Murphy-Hill et al. \citep{murphy2014cowboys} conducted a survey of 364 developers on how game development differs from other software development. Their results show that several strands of software engineering research and active practices can have a significant impact on game development. Smaller demo games may be a good starting point for novices to get into game development.

Software prototyping is a practice that is actively used in the domain of software development to reduce the number of design iterations required \citep{8321218}. Prototypes in commercial settings are helpful in obtaining a better idea about the requirements of a client, getting customer feedback and mitigating the risks involved with product development. 

Prototyping is an important step in the game development process that allows exploration and experimentation during the learning stages. Rapid prototyping is an approach to prototyping that emphasizes the rapid development of ideas into functional prototypes to demonstrate or check the feasibility of a concept. This is usually an iterative approach and is extensively used in software development and hardware design \citep{muller2021spatialproto,10.1145/3459926.3464757}. In software development, a prototype is considered a simulation or a model that allows one to experience the system \citep{warfel2009prototyping}. In terms of game development, prototypes can take several forms, such as storyboards and sketches \citep{manker2011game}. In this paper, we use the word prototypes to refer to small playable versions of a game that may be used to demonstrate a key game mechanic. 

Rapid prototyping is often used in game design to test game concepts and accompany game pitches \citep{bomstrom2020synchronizing}. Game-based learning and the introduction of game design into course content is another venue where prototyping has been found to be a crucial step \citep{zhang2021using, pirker2016value}. Game prototyping thus being a much more hands-on process than brainstorming, helps to test and improve ideas continuously \citep{gray2010gamestorming}. Shareable prototypes could help demonstrate ideas and improve collaboration in team projects.  Further game prototyping can also help students gain important job skills like project management and effort estimation  \citep{pirker2016value} which we feel might make it a good fit for inclusion in introductory courses.

Development groups can end up wasting time having to deal with heterogeneity and balancing all aspects of a game for an idea that might not be feasible \citep{10.1145/971564.971590}. This might occur as novice game developers skip out on the prototyping stages for simplicity \citep{schaeffer2017visionary}. 
 Game engine code which deals with defining the physics of the world within a game is often specific to the engine chosen  
and can often be confusing requiring repetitive tweaking and sound algorithmic knowledge to get components to work as desired \citep{10.1145/971564.971590}. Considering the overhead involved for novice game developers to get used to a specific game engine, there may be perks to building a system that is much easier to modify for the purpose of building rapid prototypes \citep{hmeljak2020developing}. \textit{NoteG} which we propose in this paper is a computational notebook that works towards this goal and aims to simplify the process of rapid game prototyping.

Computational notebooks have emerged as good learning resources for programming tasks with hands-on demonstration capabilities \citep{rule2018exploration}. 
 Even outside of traditional software engineering backgrounds, users have found computational notebooks to be useful making it a rapidly-emerging landscape. Some examples are shareable visualizations and executable code snippets accompanied by documentation that find use across academia and industry \citep{lau2020design}.
The ability of computational notebooks to enable authoring live code and integrate step-wise output in a single document are a few of the reasons why they have become popular among developers and data scientists\citep{granger2021jupyter}. 
They also help in understanding concepts and their applications through the immediate execution of code snippets \citep{walden2013informatics}. Thus computational notebook-like environments may be used to assist prototyping by providing a simplified environment for developers and enabling sharing of live code that can be used to tweak and test parameters among collaborators.

Even though there are several works in the literature that try to simplify gamification or teach game development, our work differs in the following ways:
\begin{itemize}
  \item 
Uses a computational notebook based approach which makes it easy to test, demonstrate and collaborate on a project.
  \item Attempts to reduce the overhead involved in developing a game such as learning to use a Game Engine in situations where the focus is more on the code involved.
  \item For students with a keen interest in game development, this may help incorporate a practice of rapid prototyping.
\end{itemize}

In this paper, we demonstrate how live coding and cell-based execution can simplify game prototyping, which in turn may help novice programmers learn game development or can be used to engage students in computing courses. This approach is illustrated through a computational notebook called \textit{NoteG} geared towards prototyping simple 2D role-playing games (RPG). We also evaluate the approach with 18 volunteers with overall positive response and a net promoter score of 66.

\section{Related Work}

Game development is a domain that is continuously evolving 
and has become increasingly complex over the years \citep{marklund2019empirically}.
With the increasing technical complexity involved, there is often a high technical risk associated with game design decisions \citep{10.1145/971564.971590}. 
The increasing divide between ease of use and game development speed using modern-day game engines \citep{10.1145/3341525.3387428} can make it difficult for a beginner to develop a game. This is where Model Driven Game Development (MDGD) approaches have been used by developers to reduce development time with reusable artifacts \citep{10.1145/1541895.1541909,marchisio2020automatic,baldassarre2021phydslk}. 
One cannot predict how long it can take to build a new component or if it will be implemented in the final game. Consideration must also be given to how game components may interact with the rest of the system despite the technical feats involved \citep{10.1145/971564.971590}. This drives the need for inclusion of game prototyping in the development cycle, where essentially one builds tiny experimental games iteratively, to validate and showcase the game ideas \citep{10.1145/1810295.1810325}. Over the years, \textit{game jams} \footnote{\url{https://globalgamejam.org/what-game-jam}} have gained popularity by facilitating prototyping and in turn, increasing engagement and sparking creativity towards disruptive game mechanics.
Apart from its use in the industry, rapid game prototyping has also found use in improving the quality of games and areas of education that can make use of the development of games to teach or practice relevant concepts \citep{cooper2012towards,mashuri2021developing}.

\subsection{Prototyping in Game Development and Education}

Similar to software development, the adoption of rapid prototyping is actively involved in enhancing the game development cycle. Borg et al. describe how good game development practices have drawn from software engineering research \citep{borg2019video} and emphasize the need to explore its impact on game development depending on the constraints. Prototyping though accepted as a core practice has been reported to be adopted only by a small fraction of the teams surveyed due to time constraints and sometimes has led to use of rapid prototyping \citep{borg2019video}.
In the field of education, it has been applied in many forms such as game development electives for university students \citep{10.1145/3159450.3159588}, complementing the existing academic curriculum and for creating content that can engage students \citep{10.1145/3328778.3367011}.

 Apart from the use of educational games, the development of games has been explored as a fun way to learn systems analysis in the course designed by Nikunj Dalal in his article \citep{dalal2012teaching} where he notes the use of high-level software that is much simpler to use than conventional game engines.
 
 Simplified game engines with abstracted code may also help students get started with game design \citep{10.1145/3341525.3387428}. These have made the domain ever more accessible to students and instructors alike enabling their integration into courses and this has been reported to have an overall positive impact \citep{10.1145/2538862.2538899}. The use of such approaches may enable students to explore various game mechanics and may help in providing a better understanding of core game design elements. 

J. Blow explains that overall project size and high domain-specific requirements are two roadblocks that may cause complexity in game development \citep{10.1145/971564.971590}. 
Instead of focusing on simplifying game development, we wish to bring focus on making game prototyping more accessible to novices and enable developers to inculcate the habit of software prototyping. Perker et al. in their paper \citep{pirker2016value} express how game prototyping helps in teaching various important skills such as project management, effort estimation, networking and even communication skills  required by the industry. 

Game prototyping encourages creative exploration of the domain and allows one to select the best idea from a set of alternatives based on technical feasibility and available resources at the moment. Rapid game prototyping has been explored not only in connecting academia and industry but also in controller design to explore how it can affect the social experience of video games \citep{10.1145/2468356.2479512} with the help of Arduinos. Iris Soute et al. developed Rapido a prototyping platform for outdoor games \citep{10.1145/3105704}. They also explain how their tool enabled designers to focus on the core concepts and build prototypes without worrying about lower-level issues. 

\subsection{Use of Computational Notebooks}

The idea of computational notebooks is based on literate programming which emphasizes the understandability of the code \citep{knuth1984literate} and that program logic be written in human-readable forms such as code snippets and macros. 
With the introduction of the \textit{Jupyter Notebook} \citep{kluyver2016jupyter}, computational notebooks have become quite popular and widely used by developers. It is a readable, reproducible and executable notebook that can be used for publishing code and generating results and visualization.

There have been a number of different ways in which computational notebooks have been employed. O'Hara et al. proposed a computational notebook to teach AI and introduced a new way of teaching and learning \citep{o2015computational}. Corno et al. \citep{corno2019towards} introduced a notebook for IoT development by executing notebook on top of the \textit{Docker Engine}. Many other computational notebooks like \textit{GrapePress} for Graph Transformations\citep{weber2021grapepress} and \textit{visJS2jupyter} for visualizing mechanisms of biological processes \citep{rosenthal2018interactive} have been proposed by researchers to help developers by providing interactive and user-friendly environments. 

\subsection{Prototyping}

There have also been explorations into automation of game prototyping in 2D platform games through model-driven engineering \citep{10.1145/1541895.1541909}. Reducing the overhead involved in developing a game using a user-friendly framework in this stage may help towards promoting the adoption of game prototyping among beginners. A more code-centred approach may also make it easier to integrate into learning environments for novice game developers who are more accustomed to programming. Rapid prototyping can also help students in making dynamic changes to a game by allowing for creative risks rather than technical, through mixing and matching components \citep{pirker2016value}. The use of such game prototypes could further promote analytical thinking and help in teaching design patterns for object-oriented programming and similar topics \citep{rule2018exploration}.

Rapid prototyping has also been proposed for applications such as web / mobile app development \citep{steinglass2017app} to provide a learning environment for students to learn software engineering principles. Challenging tasks such as terrain generation which requires a large number of iterations have explored rapid prototyping and ways to support the same \citep{wang2020sketch2map} but this involves sketches and landscapes. There have also been tools that directly aim to simplify the process of creating game mechanics, Volkovas et al. have proposed the use of a language for prototyping and compare it against prevailing techniques for 2D games \citep{volkovas2019mek}. 


We see that computational notebooks possess a lot of features that can help in creating rapid prototypes and that it has previously been successfully applied to create user-friendly and shareable code. However, to the best of our knowledge, despite its wide application in a number of domains the use of computational notebooks for enhancing and simplifying game prototyping for novice game developers has not been explored in the current literature.

\section{Design and Development of \textit{NoteG}}
\begin{figure*}
  \centering
  \includegraphics[width=\textwidth]{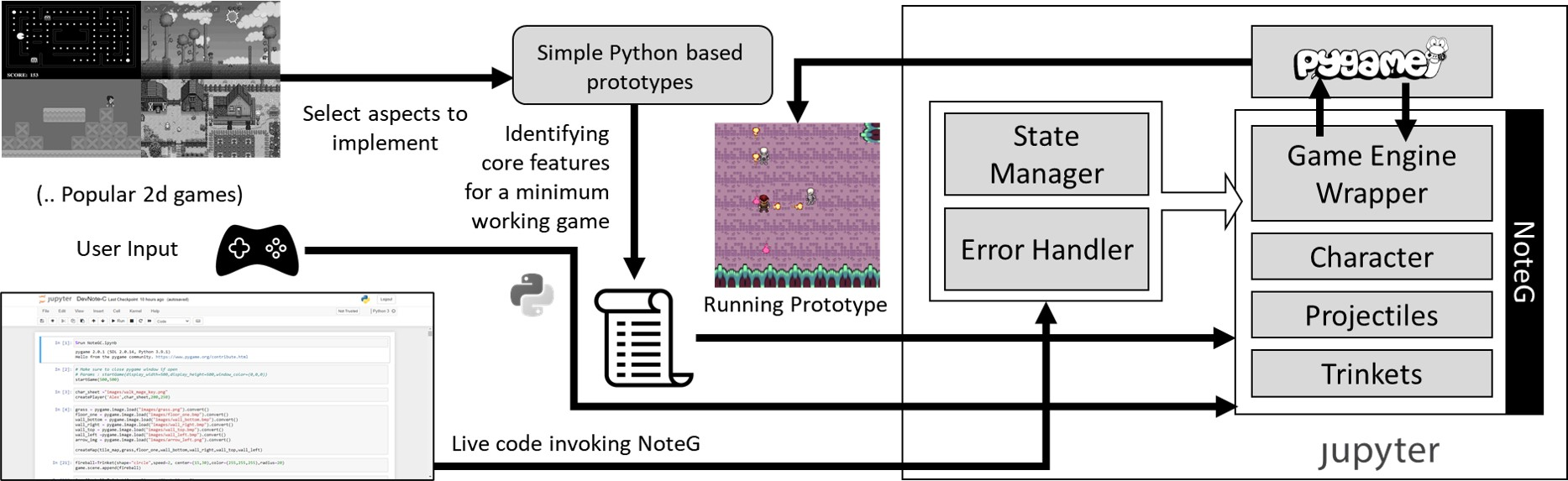}
  \caption{Workflow of building \textit{NoteG} for Rapid Game Prototyping. }
  \label{fig:1}
\end{figure*}
In this section, we describe the proposed approach, design decisions and development process of \textit{NoteG} as shown in Fig. \ref{fig:1}. The primary objectives of \textit{NoteG} are:
\begin{itemize}
  \item Obtaining a simplified command-set, useful in developing multiple games without a steep learning curve.
  \item Allow easy expansion or modification of the existing command-set for specific use cases.
  \item Use a class-based programming paradigm with support for block-based systems \citep{kolling2015frame} to allow inheritance for future game objects and also to support hot-swapping at run-time. 
  \item Incorporate error management into the state manager so that any object that causes errors is removed from the test environment providing an error traceback to prevent crashes.
  \item To support prototype development by enabling proper documentation, code folding and shareability.
\end{itemize}

Keeping in mind the aforementioned objectives and based on the feedback received during the process, we built \textit{NoteG}. It is a computational notebook that serves as a basic foundation and can be invoked to begin the prototyping process. The following sections go into more detail about the key features of the current implementation and how it can be extended based on the user's requirements.
\subsection{Background Framework}

\textit{NoteG} has been built as a \textit{Jupyter Notebook}\footnote{\url{https://jupyter.org/}}. This is because its cell-based implementation allows the developer to break down the code into more readable blocks and hence makes it easier to work with.
\textit{Jupyter Notebook} is a common and popular tool that users may have already been exposed to. Another advantage to using it is the number of extensions it supports making it easier to add functionality such as block-based programming for beginners which we go over in the next section. Even though Python is not the fastest programming language, its simplified syntax helps in making it more accessible than C\# and C++ which are used by popular game engines such as \textit{Unity
\footnote{\url{https://unity.com/}}
} and \textit{Unreal Engine\footnote{\url{https://www.unrealengine.com}}} respectively. To inherently support cross-platform development, \textit{NoteG} uses \textit{PyGame}\footnote{\url{https://www.pygame.org/}}, which is a collection of modules designed to write video games.
The top-level objects generated at run-time can be invoked later by a user to implement custom scenarios.

\subsection{Structure and Functionality}
\begin{figure*}
  \centering
  \includegraphics[width=\textwidth]{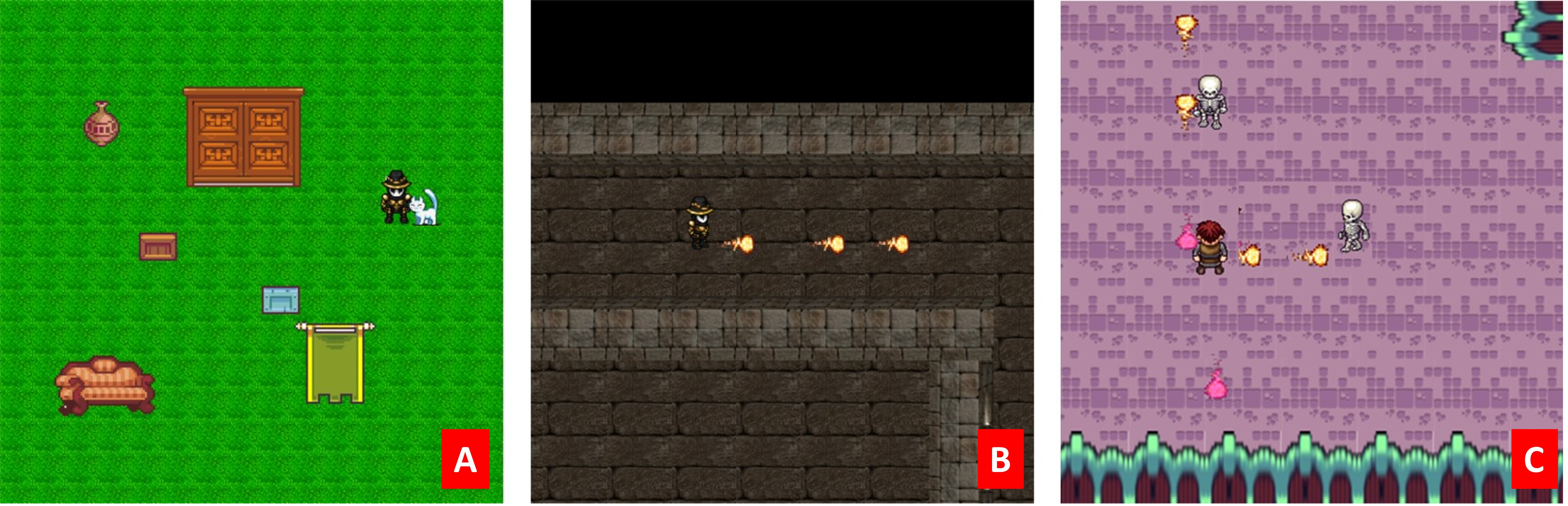}
  \caption{Sample test scenes implemented in \textit{NoteG} (a) RPG game setup (b) Dungeon with multiple rooms (c) Open world with enemies.}
  \label{fig:2}
\end{figure*}
\begin{figure}[!t]
  \centering
  \includegraphics[width=\linewidth, height=4cm]{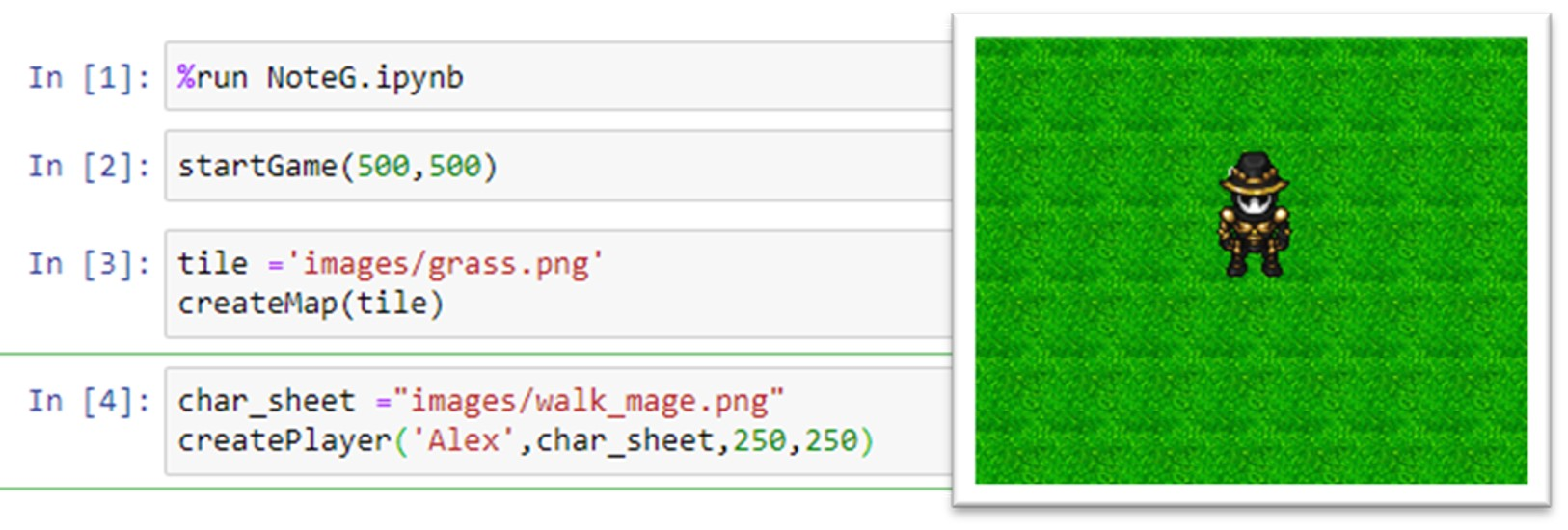}
  \caption{ Loading assets from code blocks (left side) and getting started with a scene to work on (right side).}
  \label{fig:3}
\end{figure}

\begin{table*}[!t]
  \caption{API and functions implemented in \textit{NoteG} that could be used to build a rapid game prototype.}
  \centering
  \label{tab:func}
  \begin{tabular}{|p{0.35\textwidth}|p{0.65\textwidth}|}
    \hline
    \textbf{Functions} & \textbf{Functionality} \\
    \hline
    \textit{start\_game(w,h,c)} & Starts a \textit{PyGame} instance (only one at a given time) based on the given width, height and window colour. It also implements \textit{threading} to add live support. This method handles the creation of a thread and starts the execution within it so that further execution is not blocked and other cells can be executed.\\
    \hline
    \textit{create\_map(s)} & Takes path to an image for tiling the entire map. The codebase also implements another variant of this function to facilitate the creation of custom maps, this function takes a 2D array as a tilemap and sprites for the floor and walls as parameters as shown in Fig. \ref{fig:2}(b).\\
    \hline
    \textit{create\_player(name,sprite,x,y)} & Creates a player object with keyboard controls baked in. The Character class which acts as a parent to this object also adds support for collision detection and for the addition of public variables like position and health that can be modified on run-time.\\
    \hline
    \textit{add\_trinket(trinket)} & Takes a \textit{Trinket()} class object which could be a shape, sprite or an image and adds it to the scene with background checks to handle any errors that might occur later on as it is modified.\\
    \hline
    \textit{spawn\_enemy(sprite,x,y)} & Adds an enemy based on logic implemented in Enemy() class. The update function can be redefined and parameters tweaked to test out various configurations, for example, one could run the A* algorithm on the tilemap for a better follow player logic \footnote{\url{http://theory.stanford.edu/~amitp/GameProgramming/AStarComparison.html}}. By default, the enemy follows and fires projectiles at the player.\\
    \hline
    \textit{callback\_prob(function,p)} & Defines a random callback function which is called with a probability of p with every update cycle. It can also be further extended to trigger events in the game.\\
    \hline
    \textit{game.refresh\_scene()} & Enables the user to force the screen to reset to the base state with only the player and tile-map in case an error occurs which is not handled properly.\\
  \hline
\end{tabular}
\end{table*}

In order to demonstrate how \textit{Jupyter Notebooks} can be used to enhance game prototyping, we have implemented custom classes and functions in \textit{NoteG}. Specific functionalities were selected so as to enable building a bare-bones RPG type game that can be used for testing. In order to select the functions that make up the current command-set we first built basic prototypes for a number of popular 2D games like \textit{The classic snake game} and a few other pixel art games like \textit{Terraria}. A few examples of simple scenes generated while using the notebook are shown in Fig. \ref{fig:2}. 
Table \ref{tab:func} presents the API for the game engine wrapper which was used to generate these samples and its functionality.

Further direct object manipulation which is described in detail in Section \ref{sec:func} inherently allows for hot-swapping public variables and functions of all in-game objects. 
In case modifications to an object causes errors, the object is removed from the scene with a traceback error message to prevent crashing the entire game while prototyping. This task is performed by the error handler.
A scene list keeps track of all objects in the scene and can be used in cases when reference to the object is not maintained. A number of utility functions can also be found in the codebase in addition to the ones described in Table \ref{tab:func}. These help improve the user experience and give more control over the spawned components in the game.
Using the \textit{hide()} function allows the developer to hide a cell of the notebook from view to effectively fold the code before sharing. 
Beginners can also use block-based programming extensions like jigsaw\footnote{\url{https://jupyter4edu.github.io/jupyter-edu-book/jupyter.html}} which may be invoked to get familiar with Python.
Other than these there are also a few sprite-sheet management functions to facilitate cleaning and extracting images from image files. A few custom functions were also added for illustration purposes 
and have been described in documentation cells of the notebook.

\subsection{Overview of Functions in NoteG}
\label{sec:func}
Functions in \textit{NoteG} can be used for a wide genre of 2D games Fig. \ref{fig:2} shows a few examples of these. The code block for starting a scene with loaded assets\footnote{All assets used for illustration are obtained from \url{https://opengameart.org/}} can be seen in Fig. \ref{fig:3}. We see that this contains the initial setup parameters required here the display window size is specified. Fig. \ref{fig:2}(a) shows an RPG scenario with a plain texture repeated as the ground layer using the \textit{create\_map()} function. The other two screenshots depict a dungeon (Fig. \ref{fig:2}(b)) as well as an open-world setup (Fig. \ref{fig:2}(c)) respectively both of which have bounding walls and styled tiles defined using a 2D array that is passed as an optional argument while creating the map.

Once the game is started, the other required components can be added dynamically. Fig. \ref{fig:3} shows how the \textit{create\_player()} function can create a named player object for future reference. Passing different sprite sheets can change how the character looks, helper functions auto-configure the animation system and movement system for the player character. 

Similarly, enemy characters can be added as shown in the last screenshot of Fig. \ref{fig:2}(c), using the \textit{spawn\_enemy()} function which is a modification of the player function. By default, the enemy is configured to use the A* algorithm and hunt down the player but these parameters can be edited as per the user's requirement. The object-oriented structure makes it easy to swap out any of the default logic with custom functions as shown in Fig. \ref{fig:4} and Fig. \ref{fig:5}.

Fig. \ref{fig:4} illustrates how the properties of projectiles can be altered while the game is still running. Error handlers ensure that any changes that may cause the game to crash are dealt with by terminating only the problematic objects. Fig. \ref{fig:5} shows how game mechanics that didn't exist in the base implementation of \textit{NoteG} can be easily added by a user. Even users without much programming experience can do these using helper extensions that can be enabled in the Notebook before invoking the same. Existing classes and those provided by \textit{PyGame} can also be modified as shown in Fig. \ref{fig:6} which implements collectables.

\subsection{Live Execution and Perks}

The use of computational notebooks for game prototyping allows for quick testing and cell-based execution. These cells allow one to write new code that can be executed independently. In the context of game prototyping, this allows us to add to the functionality of components that are already rendered at any point in the future. Further, notebooks also make it easy to document the code and make it readable to other developers.

\begin{figure}[!t]
  \centering
  \includegraphics[width=\linewidth]{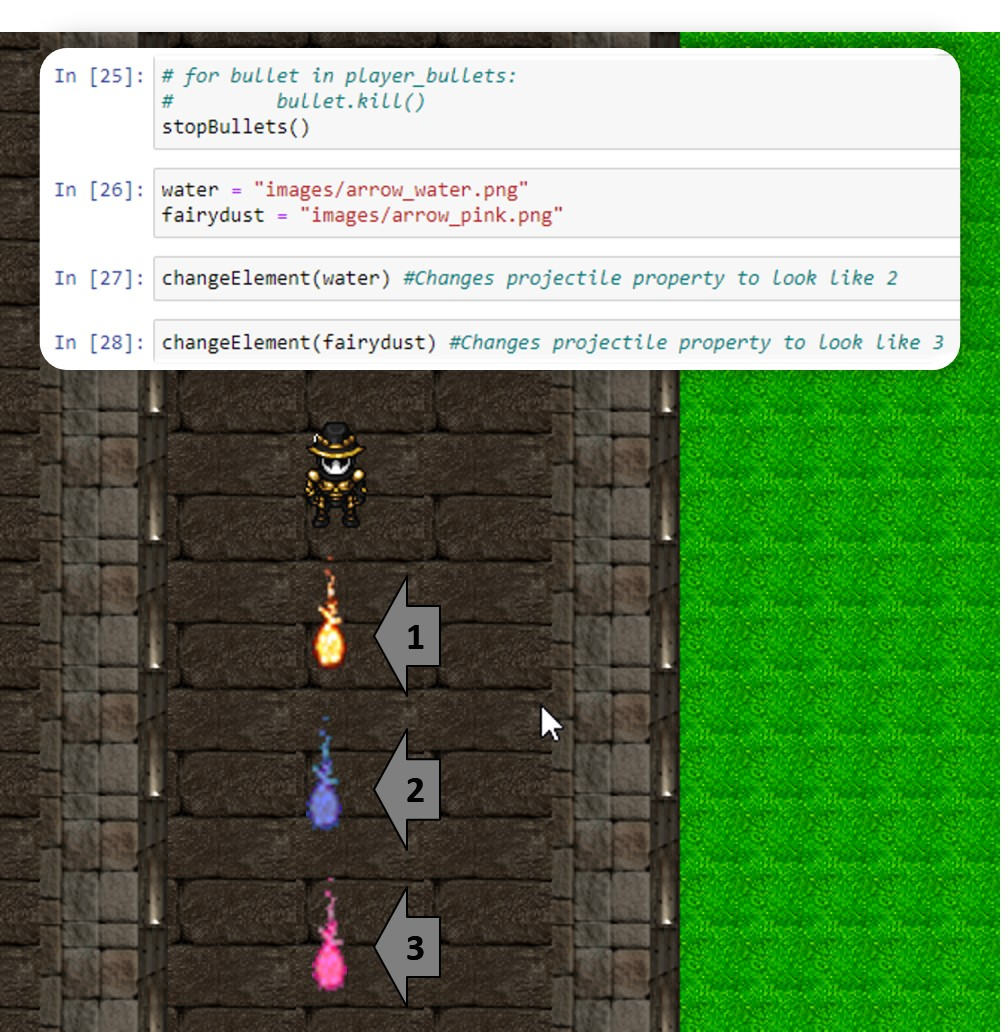}
  \caption{Hot swapping properties of the bullet object to change from state 1 to 3 using \textit{NoteG} while the game is running to demonstrate the use of live code in-game prototypes.}
  \label{fig:4}
\end{figure}
\begin{figure}[!t]
  \centering
  \includegraphics[width=\linewidth]{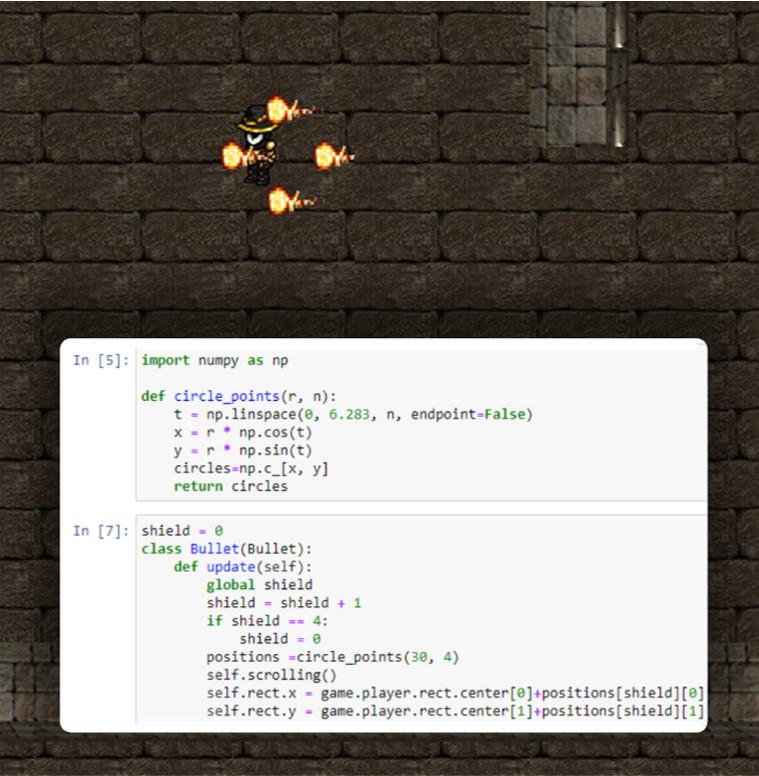}
  \caption{Tweaking behaviour by updating member functions to form a projectile shield as implemented by a volunteer in the user survey.}
  \label{fig:5}
\end{figure}

Even though features like hot-swapping which allows us to replace code without recompiling the full project already exist in popular game engines such as Unity. However many of the novice game developers who we interacted with were not aware of them which became apparent from our user survey (Section \ref{sec:eval}). An example of this is shown in Fig. \ref{fig:4}. These also require the user to follow a particular paradigm so that their code can support the feature. 

Getting used to such workflows which are directly available due to the class-based background implementations in \textit{NoteG} may help make novice game developers aware and adopt these design practices when they switch to the development stage. Fig. \ref{fig:5} depicts one of the run-time modifications on the projectile class created by a volunteer in the user survey.

\subsection{Modularity and Shareability}

\begin{figure}[!t]
  \centering
  \includegraphics[width=\linewidth]{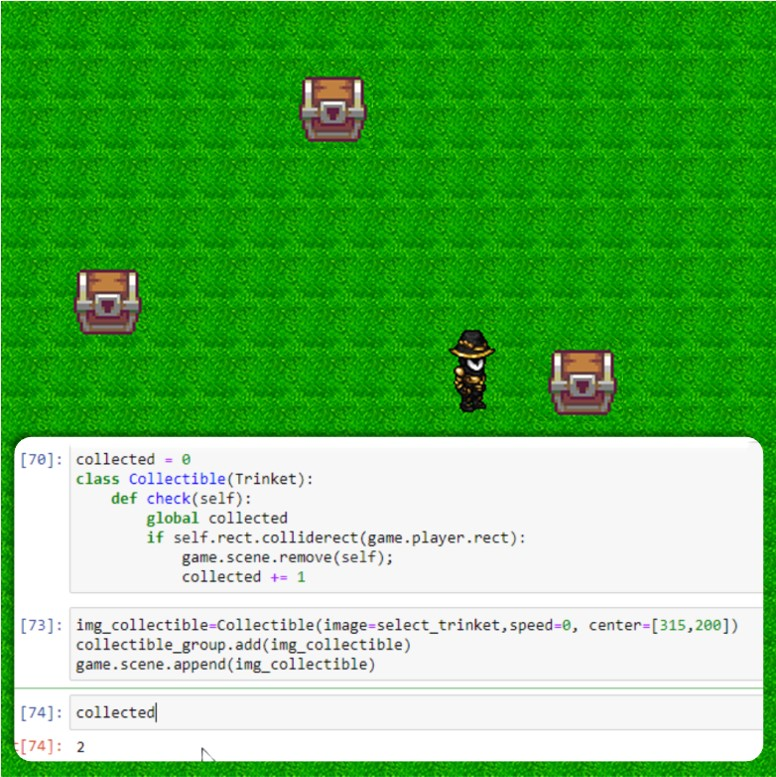}
  \caption{Adding a collectable type object by extending the Trinket class to demonstrate modularity and scope of expansion.}
  \label{fig:6}
\end{figure}
The current design of \textit{NoteG} inherently introduces classes, objects and functions and how they can be used to build reusable components. Since \textit{NoteG} is written as a wrapper on \textit{PyGame}, developers can easily make and append modules to \textit{NoteG} for implementing features that they commonly use. Some examples are splash screens and collectable objects which can be created as per the developer's requirement. An example of how the Trinket class in \textit{NoteG} can be extended to work as an in-game collectable at run-time with minimal code is shown in Fig. \ref{fig:6}. New components can also be added programmatically while testing, like the "enemy skeletons"  as shown in Fig. \ref{fig:2}(c).

Another feature that has led to the growth in popularity of notebooks in a wide variety of domains is the ease of shareability of live code \citep{wang2020better}. Collaboration is a key part of the game development cycle and can significantly impact the direction of the game \citep{whitson2018voodoo}.
Notebooks make it much easier than the traditional practices to share the prototypes with others who might not be very technically inclined but would like to retain the ability to test various situations in the game.

Game development is often a collaborative project. For this reason, game prototypes often have to be shared with other co-developers and reviewers. While sharing, packing game features as live code within a notebook can enable easy tweaking without any additional effort from the developer. This is essential
in ensuring the simplicity of the review process.

\section{Evaluation}
\label{sec:eval}
In line with the literature currently available,
\textit{NoteG} has also been evaluated through a questionnaire-based user survey  \citep{10.1145/2592235.2592238,nunez2017model}. The survey was targeted at novice game developers most of whom were at the high school and undergraduate levels. It should be noted that the participants were novice Python programmers with less than six months of experience in game development. 
It consisted of 11 questions 4 of which were aimed at gathering demographic information of the participants and the rest of the questions were 5 point Likert scale-based and aimed at understanding the perception of \textit{NoteG} among the volunteers. 
The questionnaire asked volunteers about their awareness of hot-swapping like features in other game engines and their experience with game development and prototyping. We also enquired if they felt that using \textit{NoteG} made game prototyping faster and collected feedback on implemented features. The full questionnaire and survey data can be found in the footnote\footnote{Survey Data: \url{https://dataosf.page.link/NoteG}}.


%

\begin{figure}[!t]
  \centering
\includegraphics[width=\linewidth]{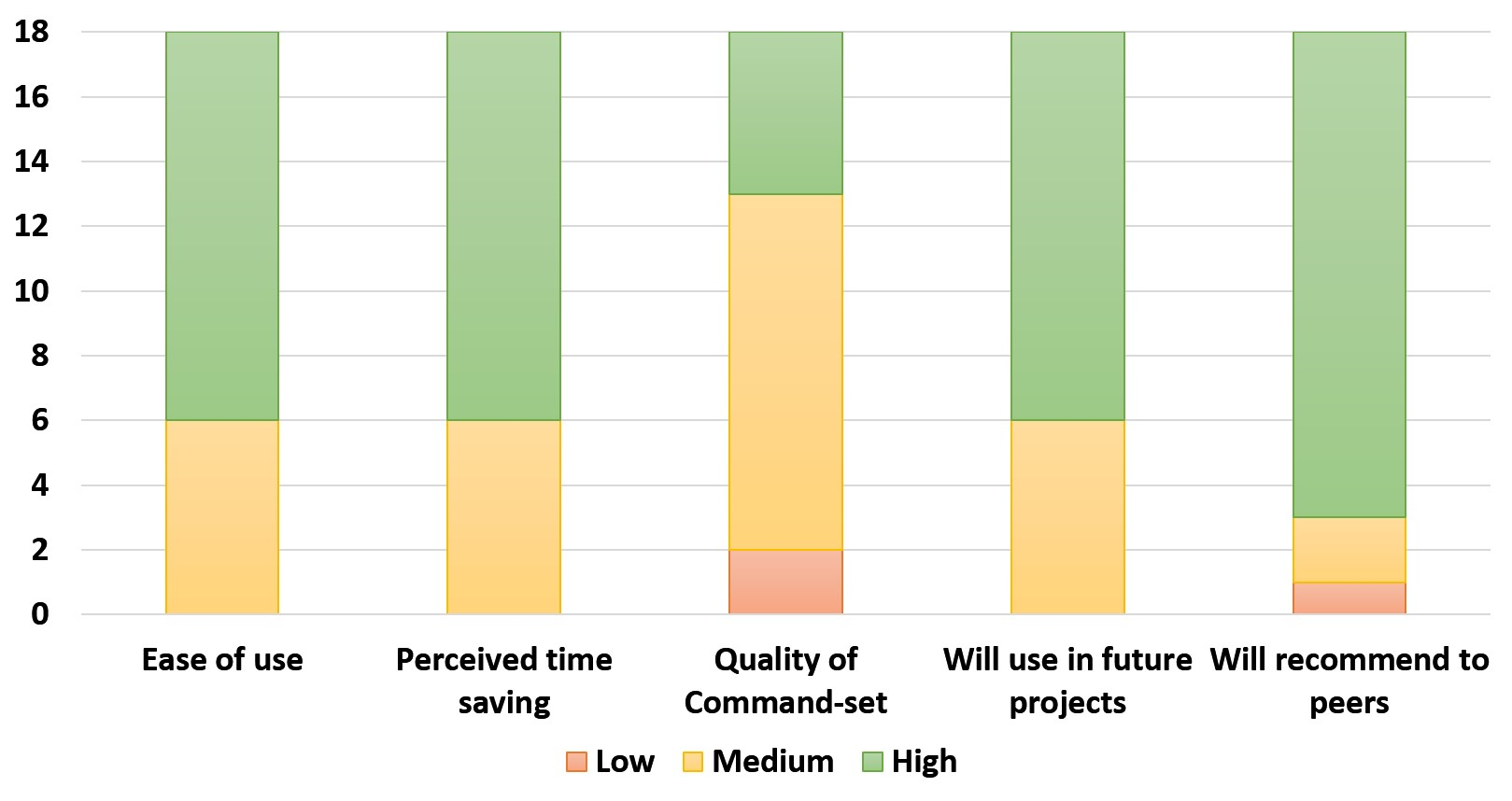}
  \caption{Results of user survey: the 5-point Likert Scale scores were converted into 3 brackets (High, Medium and Low).}
  \label{fig:7}
\end{figure}

The majority of the participants (55\%) belonged to the 20-29 age group while the rest are 15-19 years old. Out of the 18 volunteers, 66.7\% are male and 33.3\% are female. 50\% of the participants had prior experience with building games, whereas a majority of the rest are in the process of learning game development. Out of these, only 33.4\% had exposure to game prototyping. All the volunteers were given detailed documentation of the command-set and explained about the working of \textit{NoteG}, a tutorial on setting up \textit{Jupyter Notebook} and getting started with the tool is also provided. Participants were asked to try out the tool on their own by building simple prototypes and sharing them with us. The 5-point Likert scale data from the user survey was recorded as shown in Fig. \ref{fig:7}. Correlation analysis was conducted and since the Likert scale data is ordinal, we used Spearman correlation for finding the required correlation coefficients.

Of the people who responded to the survey around 60\% had prior exposure to game development while only 55\% had a basic or decent understanding of game prototyping. From the survey responses, we found a significant moderate positive correlation (level of significance = 0.01) between a person’s level of exposure to game development and his/her level of understanding of game prototyping. Also, such a person is more aware of hot swapping and this correlation was positive and significant at a 0.05 level of significance. 66\% of the respondents considered game prototyping using \textit{NoteG} to be easy or very easy and 50\% of those who are learning game development consider using \textit{NoteG} for prototyping to be easy. Most of the respondents of the survey were of the opinion that using \textit{NoteG} makes game prototyping less time-consuming. 

55\% of the respondents expressed that they are likely to use such computational notebooks in the future. Further, a majority of the users who have developed games before stated that they would use computational notebooks for game prototyping in their future projects. When it comes to recommending \textit{NoteG} the response was positive with a net promoter score of 66 (assuming those responding as “moderately likely” and “highly likely” as promoters and the rest as detractors). It was observed that 50\% or more respondents who are learning game development are highly likely to recommend \textit{NoteG} to their peers.

\section{Discussion and Future Work}
 
The computational notebook \textit{NoteG} is developed as an example of how such notebooks can be used for game prototyping which has not been explored in the current literature 
Another way to improve the current implementation would be by supporting a wider variety of games and by adding embeddable workflows that can help users to streamline the prototyping process through direct integrations with existing game engines. 

Considering the popularity of computational notebooks among data scientists and machine learning practitioners, we plan on extending the scope of our work taking inspiration from toolkits like \textit{Gym}\footnote{\url{https://gym.openai.com/docs/}} which helps to develop and compare ML algorithms. The ability of agents to interact with environments in \textit{Gym} could be exploited to enable easy training and testing of Non-Player Character AI's within the prototyping tool itself. This would take the tool closer to our ideal of enabling upcoming developers to dynamically introduce and test game object interactions at run-time with ease.

Python is well suited for making notebooks beginner-friendly however it is not the fastest programming language.
This can become a limitation for complicated games that might require more resources and faster compilation. There is also scope for modifying the user interface of \textit{NoteG} and the implemented command set. The flexibility offered in this regard could be used to tune the tool for specific use cases. 

The evaluation could be further extended by including a wider demographic with varied levels of experience and performing a controlled study on the benefits of using this approach over other options for beginners. The reasons for which Notebooks became popular in the domain of data science could still hold true for game development and help make game development more accessible to practitioners in other domains. Future research would be required to improve upon the ways in which computational notebooks can aid game prototyping in more complex development workflows.

\section{Conclusion}
In this paper, we use computational notebooks as a tool to facilitate rapid game prototyping and propose \textit{NoteG} for prototyping 2D games. It is an interactive run-time based environment where developers can edit and run code snippets to immediately observe the changes towards testing and tweaking game parameters. We explore how \textit{NoteG} can be used to simplify the process of prototyping and simplify sharing live game code without a steep learning curve. This could be particularly useful in situations where game prototyping is used to teach computer science concepts and also to novice game developers who wish to learn to code. \textit{NoteG} has cross-platform compatibility and supports hot swapping for variables, assets and in-game behaviours of game objects created through its API. 

We plan to extend \textit{NoteG} to support play-testing and prototyping of a larger variety of games in future iterations. This approach towards game prototyping aims to make it appealing to novice game developers, while also giving the user creative freedom and an ability to explore without worrying about the underlying technicalities and rather encouraging focus on the ideas and feasibility. 
The use of computational notebooks for rapid prototyping thus has its own merits and is an area that should be explored and tested further in different development environments.

\section*{Acknowledgment(s)}
The authors would like to thank the student volunteers and those who shared the prototypes they developed using \textit{NoteG} for their time and feedback, which was used to improve the base implementation.

\section*{Disclosure statement}

No potential conflict of interest was reported by the authors.



\section*{Notes on contributor(s)}

\textbf{Noble Saji Mathews} is a B.Tech student in the Department of Chemical Engineering at Indian Institute of Technology Tirupati. He has a keen interest in software engineering research and development dealing with Education Technology and Human Computer Interaction. His projects mostly deal with tools and visualizations that aid novice developers or seek to address key issues.

\noindent
\textbf{Sridhar Chimalakonda} is an Assistant Professor in the Department of Computer Science \& Engineering at IIT Tirupati, India. He received his PhD and MS by Research in Computer Science \& Engineering from International Institute of Information Technology - Hyderabad, India. He leads the Research in Intelligent Software and Human Analytics (RISHA) Lab which primarily works in the area of Software Engineering, and specifically towards empirically and qualitatively assessing quality, reuse, architecture and evolution of a broad range of software systems (such as mobile, web, games and so on). He is passionate about the massive potential of technology for improving the quality of education and automation in educational technologies.

\bibliographystyle{apacite}
\bibliography{interactapasample}

\begin{thebibliography}{}

\bibitem [\protect \citeauthoryear {%
Aycock%
, Pitout%
\BCBL {}\ \BBA {} Storteboom%
}{%
Aycock%
\ \protect \BOthers {.}}{%
{\protect \APACyear {2015}}%
}]{%
10.1145/2729094.2742590}
\APACinsertmetastar {%
10.1145/2729094.2742590}%
\begin{APACrefauthors}%
Aycock, J.%
, Pitout, E.%
\BCBL {}\ \BBA {} Storteboom, S.%
\end{APACrefauthors}%
\unskip\
\newblock
\APACrefYearMonthDay{2015}{}{}.
\newblock
{\BBOQ}\APACrefatitle {A Game Engine in Pure Python for CS1: Design,
  Experience, and Limits} {A game engine in pure python for cs1: Design,
  experience, and limits}.{\BBCQ}
\newblock
\BIn{} \APACrefbtitle {Proceedings of the 2015 ACM Conference on Innovation and
  Technology in Computer Science Education} {Proceedings of the 2015 acm
  conference on innovation and technology in computer science education}\
  (\BPG~93–98).
\newblock
\APACaddressPublisher{New York, NY, USA}{Association for Computing Machinery}.
\newblock
\begin{APACrefURL} \url{https://doi.org/10.1145/2729094.2742590}
  \end{APACrefURL}
\newblock
\begin{APACrefDOI} \doi{10.1145/2729094.2742590} \end{APACrefDOI}
\PrintBackRefs{\CurrentBib}

\bibitem [\protect \citeauthoryear {%
Baldassarre%
\ \protect \BOthers {.}}{%
Baldassarre%
\ \protect \BOthers {.}}{%
{\protect \APACyear {2021}}%
}]{%
baldassarre2021phydslk}
\APACinsertmetastar {%
baldassarre2021phydslk}%
\begin{APACrefauthors}%
Baldassarre, M\BPBI T.%
, Caivano, D.%
, Romano, S.%
, Cagnetta, F.%
, Fernandez-Cervantes, V.%
\BCBL {}\ \BBA {} Stroulia, E.%
\end{APACrefauthors}%
\unskip\
\newblock
\APACrefYearMonthDay{2021}{}{}.
\newblock
{\BBOQ}\APACrefatitle {PhyDSLK: a model-driven framework for generating
  exergames} {Phydslk: a model-driven framework for generating
  exergames}.{\BBCQ}
\newblock
\APACjournalVolNumPages{Multimedia Tools and Applications}{}{}{1--25}.
\PrintBackRefs{\CurrentBib}

\bibitem [\protect \citeauthoryear {%
Blow%
}{%
Blow%
}{%
{\protect \APACyear {2004}}%
}]{%
10.1145/971564.971590}
\APACinsertmetastar {%
10.1145/971564.971590}%
\begin{APACrefauthors}%
Blow, J.%
\end{APACrefauthors}%
\unskip\
\newblock
\APACrefYearMonthDay{2004}{{\APACmonth{02}}}{}.
\newblock
{\BBOQ}\APACrefatitle {Game Development: Harder Than You Think: Ten or Twenty
  Years Ago It Was All Fun and Games. Now It’s Blood, Sweat, and Code.} {Game
  development: Harder than you think: Ten or twenty years ago it was all fun
  and games. now it’s blood, sweat, and code.}{\BBCQ}
\newblock
\APACjournalVolNumPages{Queue}{1}{10}{28–37}.
\newblock
\begin{APACrefURL} \url{https://doi.org/10.1145/971564.971590} \end{APACrefURL}
\newblock
\begin{APACrefDOI} \doi{10.1145/971564.971590} \end{APACrefDOI}
\PrintBackRefs{\CurrentBib}

\bibitem [\protect \citeauthoryear {%
Bomstr{\"o}m%
, Kelanti%
, Lappalainen%
, Annanper{\"a}%
\BCBL {}\ \BBA {} Liukkunen%
}{%
Bomstr{\"o}m%
\ \protect \BOthers {.}}{%
{\protect \APACyear {2020}}%
}]{%
bomstrom2020synchronizing}
\APACinsertmetastar {%
bomstrom2020synchronizing}%
\begin{APACrefauthors}%
Bomstr{\"o}m, H.%
, Kelanti, M.%
, Lappalainen, J.%
, Annanper{\"a}, E.%
\BCBL {}\ \BBA {} Liukkunen, K.%
\end{APACrefauthors}%
\unskip\
\newblock
\APACrefYearMonthDay{2020}{}{}.
\newblock
{\BBOQ}\APACrefatitle {Synchronizing Game and AI Design in PCG-Based Game
  Prototypes} {Synchronizing game and ai design in pcg-based game
  prototypes}.{\BBCQ}
\newblock
\BIn{} \APACrefbtitle {International Conference on the Foundations of Digital
  Games} {International conference on the foundations of digital games}\
  (\BPGS\ 1--8).
\PrintBackRefs{\CurrentBib}

\bibitem [\protect \citeauthoryear {%
Borg%
, Garousi%
, Mahmoud%
, Olsson%
\BCBL {}\ \BBA {} St{\aa}lberg%
}{%
Borg%
\ \protect \BOthers {.}}{%
{\protect \APACyear {2019}}%
}]{%
borg2019video}
\APACinsertmetastar {%
borg2019video}%
\begin{APACrefauthors}%
Borg, M.%
, Garousi, V.%
, Mahmoud, A.%
, Olsson, T.%
\BCBL {}\ \BBA {} St{\aa}lberg, O.%
\end{APACrefauthors}%
\unskip\
\newblock
\APACrefYearMonthDay{2019}{}{}.
\newblock
{\BBOQ}\APACrefatitle {Video game development in a rush: A survey of the global
  game jam participants} {Video game development in a rush: A survey of the
  global game jam participants}.{\BBCQ}
\newblock
\APACjournalVolNumPages{IEEE Transactions on Games}{12}{3}{246--259}.
\PrintBackRefs{\CurrentBib}

\bibitem [\protect \citeauthoryear {%
Chatham%
, Walmink%
\BCBL {}\ \BBA {} Mueller%
}{%
Chatham%
\ \protect \BOthers {.}}{%
{\protect \APACyear {2013}}%
}]{%
10.1145/2468356.2479512}
\APACinsertmetastar {%
10.1145/2468356.2479512}%
\begin{APACrefauthors}%
Chatham, A.%
, Walmink, W.%
\BCBL {}\ \BBA {} Mueller, F.%
\end{APACrefauthors}%
\unskip\
\newblock
\APACrefYearMonthDay{2013}{}{}.
\newblock
{\BBOQ}\APACrefatitle {UnoJoy! A Library for Rapid Video Game Prototyping Using
  Arduino} {Unojoy! a library for rapid video game prototyping using
  arduino}.{\BBCQ}
\newblock
\BIn{} \APACrefbtitle {CHI '13 Extended Abstracts on Human Factors in Computing
  Systems} {Chi '13 extended abstracts on human factors in computing systems}\
  (\BPG~2787–2788).
\newblock
\APACaddressPublisher{New York, NY, USA}{Association for Computing Machinery}.
\newblock
\begin{APACrefURL} \url{https://doi.org/10.1145/2468356.2479512}
  \end{APACrefURL}
\newblock
\begin{APACrefDOI} \doi{10.1145/2468356.2479512} \end{APACrefDOI}
\PrintBackRefs{\CurrentBib}

\bibitem [\protect \citeauthoryear {%
Chu%
\ \BBA {} Hung%
}{%
Chu%
\ \BBA {} Hung%
}{%
{\protect \APACyear {2015}}%
}]{%
chu2015effects}
\APACinsertmetastar {%
chu2015effects}%
\begin{APACrefauthors}%
Chu, H\BHBI C.%
\BCBT {}\ \BBA {} Hung, C\BHBI M.%
\end{APACrefauthors}%
\unskip\
\newblock
\APACrefYearMonthDay{2015}{}{}.
\newblock
{\BBOQ}\APACrefatitle {Effects of the digital game-development approach on
  elementary school Students' learning motivation, problem solving, and
  learning achievement} {Effects of the digital game-development approach on
  elementary school students' learning motivation, problem solving, and
  learning achievement}.{\BBCQ}
\newblock
\APACjournalVolNumPages{International Journal of Distance Education
  Technologies (IJDET)}{13}{1}{87--102}.
\PrintBackRefs{\CurrentBib}

\bibitem [\protect \citeauthoryear {%
Cooper%
\ \BBA {} Longstreet%
}{%
Cooper%
\ \BBA {} Longstreet%
}{%
{\protect \APACyear {2012}}%
}]{%
cooper2012towards}
\APACinsertmetastar {%
cooper2012towards}%
\begin{APACrefauthors}%
Cooper, K\BPBI M.%
\BCBT {}\ \BBA {} Longstreet, C\BPBI S.%
\end{APACrefauthors}%
\unskip\
\newblock
\APACrefYearMonthDay{2012}{}{}.
\newblock
{\BBOQ}\APACrefatitle {Towards model-driven game engineering for serious
  educational games: Tailored use cases for game requirements} {Towards
  model-driven game engineering for serious educational games: Tailored use
  cases for game requirements}.{\BBCQ}
\newblock
\BIn{} \APACrefbtitle {2012 17th International Conference on Computer Games
  (CGAMES)} {2012 17th international conference on computer games (cgames)}\
  (\BPGS\ 208--212).
\PrintBackRefs{\CurrentBib}

\bibitem [\protect \citeauthoryear {%
Corno%
, De~Russis%
\BCBL {}\ \BBA {} S{\'a}enz%
}{%
Corno%
\ \protect \BOthers {.}}{%
{\protect \APACyear {2019}}%
}]{%
corno2019towards}
\APACinsertmetastar {%
corno2019towards}%
\begin{APACrefauthors}%
Corno, F.%
, De~Russis, L.%
\BCBL {}\ \BBA {} S{\'a}enz, J\BPBI P.%
\end{APACrefauthors}%
\unskip\
\newblock
\APACrefYearMonthDay{2019}{}{}.
\newblock
{\BBOQ}\APACrefatitle {Towards Computational Notebooks for IoT Development}
  {Towards computational notebooks for iot development}.{\BBCQ}
\newblock
\BIn{} \APACrefbtitle {Extended Abstracts of the 2019 CHI Conference on Human
  Factors in Computing Systems} {Extended abstracts of the 2019 chi conference
  on human factors in computing systems}\ (\BPG~LBW0154).
\PrintBackRefs{\CurrentBib}

\bibitem [\protect \citeauthoryear {%
Dalal%
}{%
Dalal%
}{%
{\protect \APACyear {2012}}%
}]{%
dalal2012teaching}
\APACinsertmetastar {%
dalal2012teaching}%
\begin{APACrefauthors}%
Dalal, N.%
\end{APACrefauthors}%
\unskip\
\newblock
\APACrefYearMonthDay{2012}{}{}.
\newblock
{\BBOQ}\APACrefatitle {Teaching tip: Using rapid game prototyping for exploring
  requirements discovery and modeling} {Teaching tip: Using rapid game
  prototyping for exploring requirements discovery and modeling}.{\BBCQ}
\newblock
\APACjournalVolNumPages{Journal of Information Systems Education}{23}{4}{129}.
\PrintBackRefs{\CurrentBib}

\bibitem [\protect \citeauthoryear {%
{Devadiga}%
}{%
{Devadiga}%
}{%
{\protect \APACyear {2017}}%
}]{%
8321218}
\APACinsertmetastar {%
8321218}%
\begin{APACrefauthors}%
{Devadiga}, N\BPBI M.%
\end{APACrefauthors}%
\unskip\
\newblock
\APACrefYearMonthDay{2017}{}{}.
\newblock
{\BBOQ}\APACrefatitle {Tailoring architecture centric design method with rapid
  prototyping} {Tailoring architecture centric design method with rapid
  prototyping}.{\BBCQ}
\newblock
\BIn{} \APACrefbtitle {2017 2nd International Conference on Communication and
  Electronics Systems (ICCES)} {2017 2nd international conference on
  communication and electronics systems (icces)}\ (\BPG~924-930).
\newblock
\begin{APACrefDOI} \doi{10.1109/CESYS.2017.8321218} \end{APACrefDOI}
\PrintBackRefs{\CurrentBib}

\bibitem [\protect \citeauthoryear {%
Granger%
\ \BBA {} P{\'e}rez%
}{%
Granger%
\ \BBA {} P{\'e}rez%
}{%
{\protect \APACyear {2021}}%
}]{%
granger2021jupyter}
\APACinsertmetastar {%
granger2021jupyter}%
\begin{APACrefauthors}%
Granger, B.%
\BCBT {}\ \BBA {} P{\'e}rez, F.%
\end{APACrefauthors}%
\unskip\
\newblock
\APACrefYearMonthDay{2021}{}{}.
\newblock
{\BBOQ}\APACrefatitle {Jupyter: Thinking and storytelling with code and data}
  {Jupyter: Thinking and storytelling with code and data}.{\BBCQ}
\newblock
\APACjournalVolNumPages{Authorea Preprints}{}{}{}.
\PrintBackRefs{\CurrentBib}

\bibitem [\protect \citeauthoryear {%
Gray%
, Brown%
\BCBL {}\ \BBA {} Macanufo%
}{%
Gray%
\ \protect \BOthers {.}}{%
{\protect \APACyear {2010}}%
}]{%
gray2010gamestorming}
\APACinsertmetastar {%
gray2010gamestorming}%
\begin{APACrefauthors}%
Gray, D.%
, Brown, S.%
\BCBL {}\ \BBA {} Macanufo, J.%
\end{APACrefauthors}%
\unskip\
\newblock
\APACrefYear{2010}.
\newblock
\APACrefbtitle {Gamestorming: A playbook for innovators, rulebreakers, and
  changemakers} {Gamestorming: A playbook for innovators, rulebreakers, and
  changemakers}.
\newblock
\APACaddressPublisher{}{" O'Reilly Media, Inc."}.
\PrintBackRefs{\CurrentBib}

\bibitem [\protect \citeauthoryear {%
G\"{u}nther%
, M\"{u}ller%
, H\"{u}bner%
, M\"{u}hlh\"{a}user%
\BCBL {}\ \BBA {} Matviienko%
}{%
G\"{u}nther%
\ \protect \BOthers {.}}{%
{\protect \APACyear {2021}}%
}]{%
10.1145/3459926.3464757}
\APACinsertmetastar {%
10.1145/3459926.3464757}%
\begin{APACrefauthors}%
G\"{u}nther, S.%
, M\"{u}ller, F.%
, H\"{u}bner, F.%
, M\"{u}hlh\"{a}user, M.%
\BCBL {}\ \BBA {} Matviienko, A.%
\end{APACrefauthors}%
\unskip\
\newblock
\APACrefYearMonthDay{2021}{}{}.
\newblock
{\BBOQ}\APACrefatitle {ActuBoard: An Open Rapid Prototyping Platform to
  Integrate Hardware Actuators in Remote Applications} {Actuboard: An open
  rapid prototyping platform to integrate hardware actuators in remote
  applications}.{\BBCQ}
\newblock
\BIn{} \APACrefbtitle {Companion of the 2021 ACM SIGCHI Symposium on
  Engineering Interactive Computing Systems} {Companion of the 2021 acm sigchi
  symposium on engineering interactive computing systems}\ (\BPG~70–76).
\newblock
\APACaddressPublisher{New York, NY, USA}{Association for Computing Machinery}.
\newblock
\begin{APACrefURL} \url{https://doi.org/10.1145/3459926.3464757}
  \end{APACrefURL}
\newblock
\begin{APACrefDOI} \doi{10.1145/3459926.3464757} \end{APACrefDOI}
\PrintBackRefs{\CurrentBib}

\bibitem [\protect \citeauthoryear {%
D.~Hmeljak%
\ \BBA {} Zhang%
}{%
D.~Hmeljak%
\ \BBA {} Zhang%
}{%
{\protect \APACyear {2020}}%
}]{%
hmeljak2020developing}
\APACinsertmetastar {%
hmeljak2020developing}%
\begin{APACrefauthors}%
Hmeljak, D.%
\BCBT {}\ \BBA {} Zhang, H.%
\end{APACrefauthors}%
\unskip\
\newblock
\APACrefYearMonthDay{2020}{}{}.
\newblock
{\BBOQ}\APACrefatitle {Developing a computer graphics course with a game
  development engine} {Developing a computer graphics course with a game
  development engine}.{\BBCQ}
\newblock
\BIn{} \APACrefbtitle {Proceedings of the 2020 ACM Conference on Innovation and
  Technology in Computer Science Education} {Proceedings of the 2020 acm
  conference on innovation and technology in computer science education}\
  (\BPGS\ 75--81).
\PrintBackRefs{\CurrentBib}

\bibitem [\protect \citeauthoryear {%
D\BPBI M.~Hmeljak%
\ \BBA {} Zhang%
}{%
D\BPBI M.~Hmeljak%
\ \BBA {} Zhang%
}{%
{\protect \APACyear {2020}}%
}]{%
10.1145/3341525.3387428}
\APACinsertmetastar {%
10.1145/3341525.3387428}%
\begin{APACrefauthors}%
Hmeljak, D\BPBI M.%
\BCBT {}\ \BBA {} Zhang, H.%
\end{APACrefauthors}%
\unskip\
\newblock
\APACrefYearMonthDay{2020}{}{}.
\newblock
{\BBOQ}\APACrefatitle {Developing a Computer Graphics Course with a Game
  Development Engine} {Developing a computer graphics course with a game
  development engine}.{\BBCQ}
\newblock
\BIn{} \APACrefbtitle {Proceedings of the 2020 ACM Conference on Innovation and
  Technology in Computer Science Education} {Proceedings of the 2020 acm
  conference on innovation and technology in computer science education}\
  (\BPG~75–81).
\newblock
\APACaddressPublisher{New York, NY, USA}{Association for Computing Machinery}.
\newblock
\begin{APACrefURL} \url{https://doi.org/10.1145/3341525.3387428}
  \end{APACrefURL}
\newblock
\begin{APACrefDOI} \doi{10.1145/3341525.3387428} \end{APACrefDOI}
\PrintBackRefs{\CurrentBib}

\bibitem [\protect \citeauthoryear {%
Iosup%
\ \BBA {} Epema%
}{%
Iosup%
\ \BBA {} Epema%
}{%
{\protect \APACyear {2014}}%
}]{%
10.1145/2538862.2538899}
\APACinsertmetastar {%
10.1145/2538862.2538899}%
\begin{APACrefauthors}%
Iosup, A.%
\BCBT {}\ \BBA {} Epema, D.%
\end{APACrefauthors}%
\unskip\
\newblock
\APACrefYearMonthDay{2014}{}{}.
\newblock
{\BBOQ}\APACrefatitle {An Experience Report on Using Gamification in Technical
  Higher Education} {An experience report on using gamification in technical
  higher education}.{\BBCQ}
\newblock
\BIn{} \APACrefbtitle {Proceedings of the 45th ACM Technical Symposium on
  Computer Science Education} {Proceedings of the 45th acm technical symposium
  on computer science education}\ (\BPG~27–32).
\newblock
\APACaddressPublisher{New York, NY, USA}{Association for Computing Machinery}.
\newblock
\begin{APACrefURL} \url{https://doi.org/10.1145/2538862.2538899}
  \end{APACrefURL}
\newblock
\begin{APACrefDOI} \doi{10.1145/2538862.2538899} \end{APACrefDOI}
\PrintBackRefs{\CurrentBib}

\bibitem [\protect \citeauthoryear {%
Kasahara%
, Sakamoto%
, Washizaki%
\BCBL {}\ \BBA {} Fukazawa%
}{%
Kasahara%
\ \protect \BOthers {.}}{%
{\protect \APACyear {2019}}%
}]{%
10.1145/3304221.3319792}
\APACinsertmetastar {%
10.1145/3304221.3319792}%
\begin{APACrefauthors}%
Kasahara, R.%
, Sakamoto, K.%
, Washizaki, H.%
\BCBL {}\ \BBA {} Fukazawa, Y.%
\end{APACrefauthors}%
\unskip\
\newblock
\APACrefYearMonthDay{2019}{}{}.
\newblock
{\BBOQ}\APACrefatitle {Applying Gamification to Motivate Students to Write
  High-Quality Code in Programming Assignments} {Applying gamification to
  motivate students to write high-quality code in programming
  assignments}.{\BBCQ}
\newblock
\BIn{} \APACrefbtitle {Proceedings of the 2019 ACM Conference on Innovation and
  Technology in Computer Science Education} {Proceedings of the 2019 acm
  conference on innovation and technology in computer science education}\
  (\BPG~92–98).
\newblock
\APACaddressPublisher{New York, NY, USA}{Association for Computing Machinery}.
\newblock
\begin{APACrefURL} \url{https://doi.org/10.1145/3304221.3319792}
  \end{APACrefURL}
\newblock
\begin{APACrefDOI} \doi{10.1145/3304221.3319792} \end{APACrefDOI}
\PrintBackRefs{\CurrentBib}

\bibitem [\protect \citeauthoryear {%
Kletenik%
\ \BBA {} Sturm%
}{%
Kletenik%
\ \BBA {} Sturm%
}{%
{\protect \APACyear {2018}}%
}]{%
10.1145/3159450.3159588}
\APACinsertmetastar {%
10.1145/3159450.3159588}%
\begin{APACrefauthors}%
Kletenik, D.%
\BCBT {}\ \BBA {} Sturm, D.%
\end{APACrefauthors}%
\unskip\
\newblock
\APACrefYearMonthDay{2018}{}{}.
\newblock
{\BBOQ}\APACrefatitle {Game Development with a Serious Focus} {Game development
  with a serious focus}.{\BBCQ}
\newblock
\BIn{} \APACrefbtitle {Proceedings of the 49th ACM Technical Symposium on
  Computer Science Education} {Proceedings of the 49th acm technical symposium
  on computer science education}\ (\BPG~652–657).
\newblock
\APACaddressPublisher{New York, NY, USA}{Association for Computing Machinery}.
\newblock
\begin{APACrefURL} \url{https://doi.org/10.1145/3159450.3159588}
  \end{APACrefURL}
\newblock
\begin{APACrefDOI} \doi{10.1145/3159450.3159588} \end{APACrefDOI}
\PrintBackRefs{\CurrentBib}

\bibitem [\protect \citeauthoryear {%
Kluyver%
\ \protect \BOthers {.}}{%
Kluyver%
\ \protect \BOthers {.}}{%
{\protect \APACyear {2016}}%
}]{%
kluyver2016jupyter}
\APACinsertmetastar {%
kluyver2016jupyter}%
\begin{APACrefauthors}%
Kluyver, T.%
, Ragan-Kelley, B.%
, P{\'e}rez, F.%
, Granger, B\BPBI E.%
, Bussonnier, M.%
, Frederic, J.%
\BDBL {}others%
\end{APACrefauthors}%
\unskip\
\newblock
\APACrefYearMonthDay{2016}{}{}.
\newblock
{\BBOQ}\APACrefatitle {Jupyter Notebooks-a publishing format for reproducible
  computational workflows.} {Jupyter notebooks-a publishing format for
  reproducible computational workflows.}{\BBCQ}
\newblock
\BIn{} \APACrefbtitle {ELPUB} {Elpub}\ (\BPGS\ 87--90).
\PrintBackRefs{\CurrentBib}

\bibitem [\protect \citeauthoryear {%
Knuth%
}{%
Knuth%
}{%
{\protect \APACyear {1984}}%
}]{%
knuth1984literate}
\APACinsertmetastar {%
knuth1984literate}%
\begin{APACrefauthors}%
Knuth, D\BPBI E.%
\end{APACrefauthors}%
\unskip\
\newblock
\APACrefYearMonthDay{1984}{}{}.
\newblock
{\BBOQ}\APACrefatitle {Literate programming} {Literate programming}.{\BBCQ}
\newblock
\APACjournalVolNumPages{The Computer Journal}{27}{2}{97--111}.
\PrintBackRefs{\CurrentBib}

\bibitem [\protect \citeauthoryear {%
K{\"o}lling%
, Brown%
\BCBL {}\ \BBA {} Altadmri%
}{%
K{\"o}lling%
\ \protect \BOthers {.}}{%
{\protect \APACyear {2015}}%
}]{%
kolling2015frame}
\APACinsertmetastar {%
kolling2015frame}%
\begin{APACrefauthors}%
K{\"o}lling, M.%
, Brown, N\BPBI C.%
\BCBL {}\ \BBA {} Altadmri, A.%
\end{APACrefauthors}%
\unskip\
\newblock
\APACrefYearMonthDay{2015}{}{}.
\newblock
{\BBOQ}\APACrefatitle {Frame-based editing: Easing the transition from blocks
  to text-based programming} {Frame-based editing: Easing the transition from
  blocks to text-based programming}.{\BBCQ}
\newblock
\BIn{} \APACrefbtitle {Proceedings of the Workshop in Primary and Secondary
  Computing Education} {Proceedings of the workshop in primary and secondary
  computing education}\ (\BPGS\ 29--38).
\PrintBackRefs{\CurrentBib}

\bibitem [\protect \citeauthoryear {%
Kurkovsky%
}{%
Kurkovsky%
}{%
{\protect \APACyear {2009}}%
}]{%
10.1145/1539024.1508881}
\APACinsertmetastar {%
10.1145/1539024.1508881}%
\begin{APACrefauthors}%
Kurkovsky, S.%
\end{APACrefauthors}%
\unskip\
\newblock
\APACrefYearMonthDay{2009}{{\APACmonth{03}}}{}.
\newblock
{\BBOQ}\APACrefatitle {Engaging Students through Mobile Game Development}
  {Engaging students through mobile game development}.{\BBCQ}
\newblock
\APACjournalVolNumPages{SIGCSE Bull.}{41}{1}{44–48}.
\newblock
\begin{APACrefURL} \url{https://doi.org/10.1145/1539024.1508881}
  \end{APACrefURL}
\newblock
\begin{APACrefDOI} \doi{10.1145/1539024.1508881} \end{APACrefDOI}
\PrintBackRefs{\CurrentBib}

\bibitem [\protect \citeauthoryear {%
Lau%
, Drosos%
, Markel%
\BCBL {}\ \BBA {} Guo%
}{%
Lau%
\ \protect \BOthers {.}}{%
{\protect \APACyear {2020}}%
}]{%
lau2020design}
\APACinsertmetastar {%
lau2020design}%
\begin{APACrefauthors}%
Lau, S.%
, Drosos, I.%
, Markel, J\BPBI M.%
\BCBL {}\ \BBA {} Guo, P\BPBI J.%
\end{APACrefauthors}%
\unskip\
\newblock
\APACrefYearMonthDay{2020}{}{}.
\newblock
{\BBOQ}\APACrefatitle {The Design Space of Computational Notebooks: An Analysis
  of 60 Systems in Academia and Industry} {The design space of computational
  notebooks: An analysis of 60 systems in academia and industry}.{\BBCQ}
\newblock
\BIn{} \APACrefbtitle {2020 IEEE Symposium on Visual Languages and
  Human-Centric Computing (VL/HCC)} {2020 ieee symposium on visual languages
  and human-centric computing (vl/hcc)}\ (\BPGS\ 1--11).
\PrintBackRefs{\CurrentBib}

\bibitem [\protect \citeauthoryear {%
Lo%
\ \BBA {} Hew%
}{%
Lo%
\ \BBA {} Hew%
}{%
{\protect \APACyear {2020}}%
}]{%
lo2020comparison}
\APACinsertmetastar {%
lo2020comparison}%
\begin{APACrefauthors}%
Lo, C\BPBI K.%
\BCBT {}\ \BBA {} Hew, K\BPBI F.%
\end{APACrefauthors}%
\unskip\
\newblock
\APACrefYearMonthDay{2020}{}{}.
\newblock
{\BBOQ}\APACrefatitle {A comparison of flipped learning with gamification,
  traditional learning, and online independent study: the effects on
  students’ mathematics achievement and cognitive engagement} {A comparison
  of flipped learning with gamification, traditional learning, and online
  independent study: the effects on students’ mathematics achievement and
  cognitive engagement}.{\BBCQ}
\newblock
\APACjournalVolNumPages{Interactive Learning Environments}{28}{4}{464--481}.
\PrintBackRefs{\CurrentBib}

\bibitem [\protect \citeauthoryear {%
Lu%
\ \protect \BOthers {.}}{%
Lu%
\ \protect \BOthers {.}}{%
{\protect \APACyear {2014}}%
}]{%
10.1145/2592235.2592238}
\APACinsertmetastar {%
10.1145/2592235.2592238}%
\begin{APACrefauthors}%
Lu, W.%
, Sun, C.%
, Bleeker, T.%
, You, Y.%
, Kitazawa, S.%
\BCBL {}\ \BBA {} Do, E\BPBI Y\BHBI L.%
\end{APACrefauthors}%
\unskip\
\newblock
\APACrefYearMonthDay{2014}{}{}.
\newblock
{\BBOQ}\APACrefatitle {Sensorendipity: A Real-Time Web-Enabled Smartphone
  Sensor Platform for Idea Generation and Prototyping} {Sensorendipity: A
  real-time web-enabled smartphone sensor platform for idea generation and
  prototyping}.{\BBCQ}
\newblock
\BIn{} \APACrefbtitle {Proceedings of the Second International Symposium of
  Chinese CHI} {Proceedings of the second international symposium of chinese
  chi}\ (\BPG~11–18).
\newblock
\APACaddressPublisher{New York, NY, USA}{Association for Computing Machinery}.
\newblock
\begin{APACrefURL} \url{https://doi.org/10.1145/2592235.2592238}
  \end{APACrefURL}
\newblock
\begin{APACrefDOI} \doi{10.1145/2592235.2592238} \end{APACrefDOI}
\PrintBackRefs{\CurrentBib}

\bibitem [\protect \citeauthoryear {%
Manker%
}{%
Manker%
}{%
{\protect \APACyear {2011}}%
}]{%
manker2011game}
\APACinsertmetastar {%
manker2011game}%
\begin{APACrefauthors}%
Manker, J.%
\end{APACrefauthors}%
\unskip\
\newblock
\APACrefYearMonthDay{2011}{}{}.
\newblock
{\BBOQ}\APACrefatitle {Game prototyping--The negotiation of an idea} {Game
  prototyping--the negotiation of an idea}.{\BBCQ}
\newblock
\BIn{} \APACrefbtitle {DiGRA Conference: Think, Design, Play} {Digra
  conference: Think, design, play}\ (\BPGS\ 14--17).
\PrintBackRefs{\CurrentBib}

\bibitem [\protect \citeauthoryear {%
Marchisio%
, Margaria%
\BCBL {}\ \BBA {} Sacchet%
}{%
Marchisio%
\ \protect \BOthers {.}}{%
{\protect \APACyear {2020}}%
}]{%
marchisio2020automatic}
\APACinsertmetastar {%
marchisio2020automatic}%
\begin{APACrefauthors}%
Marchisio, M.%
, Margaria, T.%
\BCBL {}\ \BBA {} Sacchet, M.%
\end{APACrefauthors}%
\unskip\
\newblock
\APACrefYearMonthDay{2020}{}{}.
\newblock
{\BBOQ}\APACrefatitle {Automatic Formative Assessment in Computer Science:
  Guidance to Model-Driven Design} {Automatic formative assessment in computer
  science: Guidance to model-driven design}.{\BBCQ}
\newblock
\BIn{} \APACrefbtitle {2020 IEEE 44th Annual Computers, Software, and
  Applications Conference (COMPSAC)} {2020 ieee 44th annual computers,
  software, and applications conference (compsac)}\ (\BPGS\ 201--206).
\PrintBackRefs{\CurrentBib}

\bibitem [\protect \citeauthoryear {%
Marklund%
, Engstr{\"o}m%
, Hellkvist%
\BCBL {}\ \BBA {} Backlund%
}{%
Marklund%
\ \protect \BOthers {.}}{%
{\protect \APACyear {2019}}%
}]{%
marklund2019empirically}
\APACinsertmetastar {%
marklund2019empirically}%
\begin{APACrefauthors}%
Marklund, B\BPBI B.%
, Engstr{\"o}m, H.%
, Hellkvist, M.%
\BCBL {}\ \BBA {} Backlund, P.%
\end{APACrefauthors}%
\unskip\
\newblock
\APACrefYearMonthDay{2019}{}{}.
\newblock
{\BBOQ}\APACrefatitle {What empirically based research tells us about game
  development} {What empirically based research tells us about game
  development}.{\BBCQ}
\newblock
\APACjournalVolNumPages{The Computer Games Journal}{8}{3-4}{179--198}.
\PrintBackRefs{\CurrentBib}

\bibitem [\protect \citeauthoryear {%
Mashuri%
\ \protect \BOthers {.}}{%
Mashuri%
\ \protect \BOthers {.}}{%
{\protect \APACyear {2021}}%
}]{%
mashuri2021developing}
\APACinsertmetastar {%
mashuri2021developing}%
\begin{APACrefauthors}%
Mashuri, C.%
\BCBT {}\ \BOthersPeriod {.}
\end{APACrefauthors}%
\unskip\
\newblock
\APACrefYearMonthDay{2021}{}{}.
\newblock
{\BBOQ}\APACrefatitle {Developing Indonesian Learning Game Applications for
  Elementary School Students Using the Prototyping Method} {Developing
  indonesian learning game applications for elementary school students using
  the prototyping method}.{\BBCQ}
\newblock
\APACjournalVolNumPages{Turkish Journal of Computer and Mathematics Education
  (TURCOMAT)}{12}{4}{918--928}.
\PrintBackRefs{\CurrentBib}

\bibitem [\protect \citeauthoryear {%
M{\"u}ller%
\ \protect \BOthers {.}}{%
M{\"u}ller%
\ \protect \BOthers {.}}{%
{\protect \APACyear {2021}}%
}]{%
muller2021spatialproto}
\APACinsertmetastar {%
muller2021spatialproto}%
\begin{APACrefauthors}%
M{\"u}ller, L.%
, Pfeuffer, K.%
, Gugenheimer, J.%
, Pfleging, B.%
, Prange, S.%
\BCBL {}\ \BBA {} Alt, F.%
\end{APACrefauthors}%
\unskip\
\newblock
\APACrefYearMonthDay{2021}{}{}.
\newblock
{\BBOQ}\APACrefatitle {SpatialProto: Exploring Real-World Motion Captures for
  Rapid Prototyping of Interactive Mixed Reality} {Spatialproto: Exploring
  real-world motion captures for rapid prototyping of interactive mixed
  reality}.{\BBCQ}
\newblock
\BIn{} \APACrefbtitle {Proceedings of the 2021 CHI Conference on Human Factors
  in Computing Systems} {Proceedings of the 2021 chi conference on human
  factors in computing systems}\ (\BPGS\ 1--13).
\PrintBackRefs{\CurrentBib}

\bibitem [\protect \citeauthoryear {%
Murphy-Hill%
, Zimmermann%
\BCBL {}\ \BBA {} Nagappan%
}{%
Murphy-Hill%
\ \protect \BOthers {.}}{%
{\protect \APACyear {2014}}%
}]{%
murphy2014cowboys}
\APACinsertmetastar {%
murphy2014cowboys}%
\begin{APACrefauthors}%
Murphy-Hill, E.%
, Zimmermann, T.%
\BCBL {}\ \BBA {} Nagappan, N.%
\end{APACrefauthors}%
\unskip\
\newblock
\APACrefYearMonthDay{2014}{}{}.
\newblock
{\BBOQ}\APACrefatitle {Cowboys, ankle sprains, and keepers of quality: How is
  video game development different from software development?} {Cowboys, ankle
  sprains, and keepers of quality: How is video game development different from
  software development?}{\BBCQ}
\newblock
\BIn{} \APACrefbtitle {Proceedings of the 36th International Conference on
  Software Engineering} {Proceedings of the 36th international conference on
  software engineering}\ (\BPGS\ 1--11).
\PrintBackRefs{\CurrentBib}

\bibitem [\protect \citeauthoryear {%
Musil%
, Schweda%
, Winkler%
\BCBL {}\ \BBA {} Biffl%
}{%
Musil%
\ \protect \BOthers {.}}{%
{\protect \APACyear {2010}}%
}]{%
10.1145/1810295.1810325}
\APACinsertmetastar {%
10.1145/1810295.1810325}%
\begin{APACrefauthors}%
Musil, J.%
, Schweda, A.%
, Winkler, D.%
\BCBL {}\ \BBA {} Biffl, S.%
\end{APACrefauthors}%
\unskip\
\newblock
\APACrefYearMonthDay{2010}{}{}.
\newblock
{\BBOQ}\APACrefatitle {Synthesized Essence: What Game Jams Teach about
  Prototyping of New Software Products} {Synthesized essence: What game jams
  teach about prototyping of new software products}.{\BBCQ}
\newblock
\BIn{} \APACrefbtitle {Proceedings of the 32nd ACM/IEEE International
  Conference on Software Engineering - Volume 2} {Proceedings of the 32nd
  acm/ieee international conference on software engineering - volume 2}\
  (\BPG~183–186).
\newblock
\APACaddressPublisher{New York, NY, USA}{Association for Computing Machinery}.
\newblock
\begin{APACrefURL} \url{https://doi.org/10.1145/1810295.1810325}
  \end{APACrefURL}
\newblock
\begin{APACrefDOI} \doi{10.1145/1810295.1810325} \end{APACrefDOI}
\PrintBackRefs{\CurrentBib}

\bibitem [\protect \citeauthoryear {%
N{\'u}{\~n}ez-Valdez%
, Garc{\'\i}a-D{\'\i}az%
, Lovelle%
, Achaerandio%
\BCBL {}\ \BBA {} Gonz{\'a}lez-Crespo%
}{%
N{\'u}{\~n}ez-Valdez%
\ \protect \BOthers {.}}{%
{\protect \APACyear {2017}}%
}]{%
nunez2017model}
\APACinsertmetastar {%
nunez2017model}%
\begin{APACrefauthors}%
N{\'u}{\~n}ez-Valdez, E\BPBI R.%
, Garc{\'\i}a-D{\'\i}az, V.%
, Lovelle, J\BPBI M\BPBI C.%
, Achaerandio, Y\BPBI S.%
\BCBL {}\ \BBA {} Gonz{\'a}lez-Crespo, R.%
\end{APACrefauthors}%
\unskip\
\newblock
\APACrefYearMonthDay{2017}{}{}.
\newblock
{\BBOQ}\APACrefatitle {A model-driven approach to generate and deploy
  videogames on multiple platforms} {A model-driven approach to generate and
  deploy videogames on multiple platforms}.{\BBCQ}
\newblock
\APACjournalVolNumPages{Journal of Ambient Intelligence and Humanized
  Computing}{8}{3}{435--447}.
\PrintBackRefs{\CurrentBib}

\bibitem [\protect \citeauthoryear {%
Obaid%
, Farooq%
\BCBL {}\ \BBA {} Abid%
}{%
Obaid%
\ \protect \BOthers {.}}{%
{\protect \APACyear {2020}}%
}]{%
obaid2020gamification}
\APACinsertmetastar {%
obaid2020gamification}%
\begin{APACrefauthors}%
Obaid, I.%
, Farooq, M\BPBI S.%
\BCBL {}\ \BBA {} Abid, A.%
\end{APACrefauthors}%
\unskip\
\newblock
\APACrefYearMonthDay{2020}{}{}.
\newblock
{\BBOQ}\APACrefatitle {Gamification for recruitment and job training: model,
  taxonomy, and challenges} {Gamification for recruitment and job training:
  model, taxonomy, and challenges}.{\BBCQ}
\newblock
\APACjournalVolNumPages{IEEE Access}{8}{}{65164--65178}.
\PrintBackRefs{\CurrentBib}

\bibitem [\protect \citeauthoryear {%
O'Hara%
, Blank%
\BCBL {}\ \BBA {} Marshall%
}{%
O'Hara%
\ \protect \BOthers {.}}{%
{\protect \APACyear {2015}}%
}]{%
o2015computational}
\APACinsertmetastar {%
o2015computational}%
\begin{APACrefauthors}%
O'Hara, K.%
, Blank, D.%
\BCBL {}\ \BBA {} Marshall, J.%
\end{APACrefauthors}%
\unskip\
\newblock
\APACrefYearMonthDay{2015}{}{}.
\newblock
{\BBOQ}\APACrefatitle {Computational notebooks for AI education} {Computational
  notebooks for ai education}.{\BBCQ}
\newblock
\BIn{} \APACrefbtitle {The Twenty-Eighth International Flairs Conference.} {The
  twenty-eighth international flairs conference.}
\PrintBackRefs{\CurrentBib}

\bibitem [\protect \citeauthoryear {%
Pirker%
, Kultima%
\BCBL {}\ \BBA {} G{\"u}tl%
}{%
Pirker%
\ \protect \BOthers {.}}{%
{\protect \APACyear {2016}}%
}]{%
pirker2016value}
\APACinsertmetastar {%
pirker2016value}%
\begin{APACrefauthors}%
Pirker, J.%
, Kultima, A.%
\BCBL {}\ \BBA {} G{\"u}tl, C.%
\end{APACrefauthors}%
\unskip\
\newblock
\APACrefYearMonthDay{2016}{}{}.
\newblock
{\BBOQ}\APACrefatitle {The value of game prototyping projects for students and
  industry} {The value of game prototyping projects for students and
  industry}.{\BBCQ}
\newblock
\BIn{} \APACrefbtitle {Proceedings of the International Conference on Game
  Jams, Hackathons, and Game Creation Events} {Proceedings of the international
  conference on game jams, hackathons, and game creation events}\ (\BPGS\
  54--57).
\PrintBackRefs{\CurrentBib}

\bibitem [\protect \citeauthoryear {%
Rahimi%
\ \BBA {} Kim%
}{%
Rahimi%
\ \BBA {} Kim%
}{%
{\protect \APACyear {2019}}%
}]{%
rahimi2019role}
\APACinsertmetastar {%
rahimi2019role}%
\begin{APACrefauthors}%
Rahimi, F\BPBI B.%
\BCBT {}\ \BBA {} Kim, B.%
\end{APACrefauthors}%
\unskip\
\newblock
\APACrefYearMonthDay{2019}{}{}.
\newblock
{\BBOQ}\APACrefatitle {The role of interest-driven participatory game design:
  considering design literacy within a technology classroom} {The role of
  interest-driven participatory game design: considering design literacy within
  a technology classroom}.{\BBCQ}
\newblock
\APACjournalVolNumPages{International Journal of Technology and Design
  Education}{29}{2}{387--404}.
\PrintBackRefs{\CurrentBib}

\bibitem [\protect \citeauthoryear {%
Reyno%
\ \BBA {} Cars\'{\i}~Cubel%
}{%
Reyno%
\ \BBA {} Cars\'{\i}~Cubel%
}{%
{\protect \APACyear {2009}}%
}]{%
10.1145/1541895.1541909}
\APACinsertmetastar {%
10.1145/1541895.1541909}%
\begin{APACrefauthors}%
Reyno, E\BPBI M.%
\BCBT {}\ \BBA {} Cars\'{\i}~Cubel, J\BPBI A.%
\end{APACrefauthors}%
\unskip\
\newblock
\APACrefYearMonthDay{2009}{{\APACmonth{06}}}{}.
\newblock
{\BBOQ}\APACrefatitle {Automatic Prototyping in Model-Driven Game Development}
  {Automatic prototyping in model-driven game development}.{\BBCQ}
\newblock
\APACjournalVolNumPages{Comput. Entertain.}{7}{2}{}.
\newblock
\begin{APACrefURL} \url{https://doi.org/10.1145/1541895.1541909}
  \end{APACrefURL}
\newblock
\begin{APACrefDOI} \doi{10.1145/1541895.1541909} \end{APACrefDOI}
\PrintBackRefs{\CurrentBib}

\bibitem [\protect \citeauthoryear {%
Richard%
\ \BBA {} Kafai%
}{%
Richard%
\ \BBA {} Kafai%
}{%
{\protect \APACyear {2015}}%
}]{%
10.1145/2771839.2771926}
\APACinsertmetastar {%
10.1145/2771839.2771926}%
\begin{APACrefauthors}%
Richard, G\BPBI T.%
\BCBT {}\ \BBA {} Kafai, Y\BPBI B.%
\end{APACrefauthors}%
\unskip\
\newblock
\APACrefYearMonthDay{2015}{}{}.
\newblock
{\BBOQ}\APACrefatitle {Making Physical and Digital Games with E-Textiles: A
  Workshop for Youth Making Responsive Wearable Games and Controllers} {Making
  physical and digital games with e-textiles: A workshop for youth making
  responsive wearable games and controllers}.{\BBCQ}
\newblock
\BIn{} \APACrefbtitle {Proceedings of the 14th International Conference on
  Interaction Design and Children} {Proceedings of the 14th international
  conference on interaction design and children}\ (\BPG~399–402).
\newblock
\APACaddressPublisher{New York, NY, USA}{Association for Computing Machinery}.
\newblock
\begin{APACrefURL} \url{https://doi.org/10.1145/2771839.2771926}
  \end{APACrefURL}
\newblock
\begin{APACrefDOI} \doi{10.1145/2771839.2771926} \end{APACrefDOI}
\PrintBackRefs{\CurrentBib}

\bibitem [\protect \citeauthoryear {%
Rosenthal%
\ \protect \BOthers {.}}{%
Rosenthal%
\ \protect \BOthers {.}}{%
{\protect \APACyear {2018}}%
}]{%
rosenthal2018interactive}
\APACinsertmetastar {%
rosenthal2018interactive}%
\begin{APACrefauthors}%
Rosenthal, S\BPBI B.%
, Len, J.%
, Webster, M.%
, Gary, A.%
, Birmingham, A.%
\BCBL {}\ \BBA {} Fisch, K\BPBI M.%
\end{APACrefauthors}%
\unskip\
\newblock
\APACrefYearMonthDay{2018}{}{}.
\newblock
{\BBOQ}\APACrefatitle {Interactive network visualization in Jupyter notebooks:
  visJS2jupyter} {Interactive network visualization in jupyter notebooks:
  visjs2jupyter}.{\BBCQ}
\newblock
\APACjournalVolNumPages{Bioinformatics}{34}{1}{126--128}.
\PrintBackRefs{\CurrentBib}

\bibitem [\protect \citeauthoryear {%
Rule%
, Tabard%
\BCBL {}\ \BBA {} Hollan%
}{%
Rule%
\ \protect \BOthers {.}}{%
{\protect \APACyear {2018}}%
}]{%
rule2018exploration}
\APACinsertmetastar {%
rule2018exploration}%
\begin{APACrefauthors}%
Rule, A.%
, Tabard, A.%
\BCBL {}\ \BBA {} Hollan, J\BPBI D.%
\end{APACrefauthors}%
\unskip\
\newblock
\APACrefYearMonthDay{2018}{}{}.
\newblock
{\BBOQ}\APACrefatitle {Exploration and explanation in computational notebooks}
  {Exploration and explanation in computational notebooks}.{\BBCQ}
\newblock
\BIn{} \APACrefbtitle {Proceedings of the 2018 CHI Conference on Human Factors
  in Computing Systems} {Proceedings of the 2018 chi conference on human
  factors in computing systems}\ (\BPG~32).
\PrintBackRefs{\CurrentBib}

\bibitem [\protect \citeauthoryear {%
Schaeffer%
\ \BBA {} Palmgren%
}{%
Schaeffer%
\ \BBA {} Palmgren%
}{%
{\protect \APACyear {2017}}%
}]{%
schaeffer2017visionary}
\APACinsertmetastar {%
schaeffer2017visionary}%
\begin{APACrefauthors}%
Schaeffer, J\BPBI A.%
\BCBT {}\ \BBA {} Palmgren, M.%
\end{APACrefauthors}%
\unskip\
\newblock
\APACrefYearMonthDay{2017}{}{}.
\newblock
{\BBOQ}\APACrefatitle {Visionary Expectations and Novice Designers--Prototyping
  in Design Education.} {Visionary expectations and novice
  designers--prototyping in design education.}{\BBCQ}
\newblock
\APACjournalVolNumPages{Design and Technology Education}{22}{1}{n1}.
\PrintBackRefs{\CurrentBib}

\bibitem [\protect \citeauthoryear {%
Schulz%
, Smaradottir%
, Prinz%
\BCBL {}\ \BBA {} Hara%
}{%
Schulz%
\ \protect \BOthers {.}}{%
{\protect \APACyear {2020}}%
}]{%
schulz2020user}
\APACinsertmetastar {%
schulz2020user}%
\begin{APACrefauthors}%
Schulz, R.%
, Smaradottir, B.%
, Prinz, A.%
\BCBL {}\ \BBA {} Hara, T.%
\end{APACrefauthors}%
\unskip\
\newblock
\APACrefYearMonthDay{2020}{}{}.
\newblock
{\BBOQ}\APACrefatitle {User-Centered Design of a Scenario-Based Serious Game:
  Game-Based Teaching of Future Healthcare} {User-centered design of a
  scenario-based serious game: Game-based teaching of future
  healthcare}.{\BBCQ}
\newblock
\APACjournalVolNumPages{IEEE Transactions on Games}{12}{4}{376--385}.
\PrintBackRefs{\CurrentBib}

\bibitem [\protect \citeauthoryear {%
Siswati%
, Prihatin%
, Damayanti%
, Nafisah%
\BCBL {}\ \protect \BOthers {.}}{%
Siswati%
\ \protect \BOthers {.}}{%
{\protect \APACyear {2021}}%
}]{%
siswati2021developing}
\APACinsertmetastar {%
siswati2021developing}%
\begin{APACrefauthors}%
Siswati, B\BPBI H.%
, Prihatin, J.%
, Damayanti, A.%
, Nafisah, L.%
\BCBL {}\ \BOthersPeriod {.}\end{APACrefauthors}%
\unskip\
\newblock
\APACrefYearMonthDay{2021}{}{}.
\newblock
{\BBOQ}\APACrefatitle {Developing gamification based biology learning materials
  for senior high school students in industrial agricultural area in jember,
  indonesia} {Developing gamification based biology learning materials for
  senior high school students in industrial agricultural area in jember,
  indonesia}.{\BBCQ}
\newblock
\BIn{} \APACrefbtitle {Journal of Physics: Conference Series} {Journal of
  physics: Conference series}\ (\BVOL\ 1839, \BPG~012021).
\PrintBackRefs{\CurrentBib}

\bibitem [\protect \citeauthoryear {%
Soute%
, Vacaretu%
, Wit%
\BCBL {}\ \BBA {} Markopoulos%
}{%
Soute%
\ \protect \BOthers {.}}{%
{\protect \APACyear {2017}}%
}]{%
10.1145/3105704}
\APACinsertmetastar {%
10.1145/3105704}%
\begin{APACrefauthors}%
Soute, I.%
, Vacaretu, T.%
, Wit, J\BPBI D.%
\BCBL {}\ \BBA {} Markopoulos, P.%
\end{APACrefauthors}%
\unskip\
\newblock
\APACrefYearMonthDay{2017}{{\APACmonth{08}}}{}.
\newblock
{\BBOQ}\APACrefatitle {Design and Evaluation of RaPIDO, A Platform for Rapid
  Prototyping of Interactive Outdoor Games} {Design and evaluation of rapido, a
  platform for rapid prototyping of interactive outdoor games}.{\BBCQ}
\newblock
\APACjournalVolNumPages{ACM Trans. Comput.-Hum. Interact.}{24}{4}{}.
\newblock
\begin{APACrefURL} \url{https://doi.org/10.1145/3105704} \end{APACrefURL}
\newblock
\begin{APACrefDOI} \doi{10.1145/3105704} \end{APACrefDOI}
\PrintBackRefs{\CurrentBib}

\bibitem [\protect \citeauthoryear {%
Steinglass%
, Franke%
\BCBL {}\ \BBA {} Filman%
}{%
Steinglass%
\ \protect \BOthers {.}}{%
{\protect \APACyear {2017}}%
}]{%
steinglass2017app}
\APACinsertmetastar {%
steinglass2017app}%
\begin{APACrefauthors}%
Steinglass, A.%
, Franke, B.%
\BCBL {}\ \BBA {} Filman, S.%
\end{APACrefauthors}%
\unskip\
\newblock
\APACrefYearMonthDay{2017}{}{}.
\newblock
{\BBOQ}\APACrefatitle {App Lab: A Powerful JavaScript IDE for Rapid Prototyping
  of Small Data-backed Web Applications} {App lab: A powerful javascript ide
  for rapid prototyping of small data-backed web applications}.{\BBCQ}
\newblock
\BIn{} \APACrefbtitle {Proceedings of the 2017 ACM SIGCSE Technical Symposium
  on Computer Science Education} {Proceedings of the 2017 acm sigcse technical
  symposium on computer science education}\ (\BPGS\ 641--642).
\PrintBackRefs{\CurrentBib}

\bibitem [\protect \citeauthoryear {%
Subramanian%
, Saule%
\BCBL {}\ \BBA {} Payton%
}{%
Subramanian%
\ \protect \BOthers {.}}{%
{\protect \APACyear {2020}}%
}]{%
10.1145/3328778.3367011}
\APACinsertmetastar {%
10.1145/3328778.3367011}%
\begin{APACrefauthors}%
Subramanian, K.%
, Saule, E.%
\BCBL {}\ \BBA {} Payton, J.%
\end{APACrefauthors}%
\unskip\
\newblock
\APACrefYearMonthDay{2020}{}{}.
\newblock
{\BBOQ}\APACrefatitle {Bringing Real-World Data, Interactive Games and
  Visualizations into Early CS Courses} {Bringing real-world data, interactive
  games and visualizations into early cs courses}.{\BBCQ}
\newblock
\BIn{} \APACrefbtitle {Proceedings of the 51st ACM Technical Symposium on
  Computer Science Education} {Proceedings of the 51st acm technical symposium
  on computer science education}\ (\BPG~1391).
\newblock
\APACaddressPublisher{New York, NY, USA}{Association for Computing Machinery}.
\newblock
\begin{APACrefURL} \url{https://doi.org/10.1145/3328778.3367011}
  \end{APACrefURL}
\newblock
\begin{APACrefDOI} \doi{10.1145/3328778.3367011} \end{APACrefDOI}
\PrintBackRefs{\CurrentBib}

\bibitem [\protect \citeauthoryear {%
Volkovas%
, Fairbank%
, Woodward%
\BCBL {}\ \BBA {} Lucas%
}{%
Volkovas%
\ \protect \BOthers {.}}{%
{\protect \APACyear {2019}}%
}]{%
volkovas2019mek}
\APACinsertmetastar {%
volkovas2019mek}%
\begin{APACrefauthors}%
Volkovas, R.%
, Fairbank, M.%
, Woodward, J\BPBI R.%
\BCBL {}\ \BBA {} Lucas, S.%
\end{APACrefauthors}%
\unskip\
\newblock
\APACrefYearMonthDay{2019}{}{}.
\newblock
{\BBOQ}\APACrefatitle {Mek: Mechanics prototyping tool for 2d tile-based
  turn-based deterministic games} {Mek: Mechanics prototyping tool for 2d
  tile-based turn-based deterministic games}.{\BBCQ}
\newblock
\BIn{} \APACrefbtitle {2019 IEEE Conference on Games (CoG)} {2019 ieee
  conference on games (cog)}\ (\BPGS\ 1--8).
\PrintBackRefs{\CurrentBib}

\bibitem [\protect \citeauthoryear {%
Walden%
, Doyle%
, Garns%
\BCBL {}\ \BBA {} Hart%
}{%
Walden%
\ \protect \BOthers {.}}{%
{\protect \APACyear {2013}}%
}]{%
walden2013informatics}
\APACinsertmetastar {%
walden2013informatics}%
\begin{APACrefauthors}%
Walden, J.%
, Doyle, M.%
, Garns, R.%
\BCBL {}\ \BBA {} Hart, Z.%
\end{APACrefauthors}%
\unskip\
\newblock
\APACrefYearMonthDay{2013}{}{}.
\newblock
{\BBOQ}\APACrefatitle {An informatics perspective on computational thinking}
  {An informatics perspective on computational thinking}.{\BBCQ}
\newblock
\BIn{} \APACrefbtitle {Proceedings of the 18th ACM conference on Innovation and
  technology in computer science education} {Proceedings of the 18th acm
  conference on innovation and technology in computer science education}\
  (\BPGS\ 4--9).
\PrintBackRefs{\CurrentBib}

\bibitem [\protect \citeauthoryear {%
J.~Wang%
, Li%
\BCBL {}\ \BBA {} Zeller%
}{%
J.~Wang%
\ \protect \BOthers {.}}{%
{\protect \APACyear {2020}}%
}]{%
wang2020better}
\APACinsertmetastar {%
wang2020better}%
\begin{APACrefauthors}%
Wang, J.%
, Li, L.%
\BCBL {}\ \BBA {} Zeller, A.%
\end{APACrefauthors}%
\unskip\
\newblock
\APACrefYearMonthDay{2020}{}{}.
\newblock
{\BBOQ}\APACrefatitle {Better code, better sharing: on the need of analyzing
  jupyter notebooks} {Better code, better sharing: on the need of analyzing
  jupyter notebooks}.{\BBCQ}
\newblock
\BIn{} \APACrefbtitle {Proceedings of the ACM/IEEE 42nd International
  Conference on Software Engineering: New Ideas and Emerging Results}
  {Proceedings of the acm/ieee 42nd international conference on software
  engineering: New ideas and emerging results}\ (\BPGS\ 53--56).
\PrintBackRefs{\CurrentBib}

\bibitem [\protect \citeauthoryear {%
T.~Wang%
\ \BBA {} Kurabayashi%
}{%
T.~Wang%
\ \BBA {} Kurabayashi%
}{%
{\protect \APACyear {2020}}%
}]{%
wang2020sketch2map}
\APACinsertmetastar {%
wang2020sketch2map}%
\begin{APACrefauthors}%
Wang, T.%
\BCBT {}\ \BBA {} Kurabayashi, S.%
\end{APACrefauthors}%
\unskip\
\newblock
\APACrefYearMonthDay{2020}{}{}.
\newblock
{\BBOQ}\APACrefatitle {Sketch2map: a game map design support system allowing
  quick hand sketch prototyping} {Sketch2map: a game map design support system
  allowing quick hand sketch prototyping}.{\BBCQ}
\newblock
\BIn{} \APACrefbtitle {2020 IEEE Conference on Games (CoG)} {2020 ieee
  conference on games (cog)}\ (\BPGS\ 596--599).
\PrintBackRefs{\CurrentBib}

\bibitem [\protect \citeauthoryear {%
Warfel%
}{%
Warfel%
}{%
{\protect \APACyear {2009}}%
}]{%
warfel2009prototyping}
\APACinsertmetastar {%
warfel2009prototyping}%
\begin{APACrefauthors}%
Warfel, T\BPBI Z.%
\end{APACrefauthors}%
\unskip\
\newblock
\APACrefYear{2009}.
\newblock
\APACrefbtitle {Prototyping: a practitioner's guide} {Prototyping: a
  practitioner's guide}.
\newblock
\APACaddressPublisher{}{Rosenfeld media}.
\PrintBackRefs{\CurrentBib}

\bibitem [\protect \citeauthoryear {%
Weber%
}{%
Weber%
}{%
{\protect \APACyear {2021}}%
}]{%
weber2021grapepress}
\APACinsertmetastar {%
weber2021grapepress}%
\begin{APACrefauthors}%
Weber, J\BPBI H.%
\end{APACrefauthors}%
\unskip\
\newblock
\APACrefYearMonthDay{2021}{}{}.
\newblock
{\BBOQ}\APACrefatitle {GrapePress-A Computational Notebook for Graph
  Transformations} {Grapepress-a computational notebook for graph
  transformations}.{\BBCQ}
\newblock
\BIn{} \APACrefbtitle {International Conference on Graph Transformation}
  {International conference on graph transformation}\ (\BPGS\ 294--302).
\PrintBackRefs{\CurrentBib}

\bibitem [\protect \citeauthoryear {%
Whitson%
}{%
Whitson%
}{%
{\protect \APACyear {2018}}%
}]{%
whitson2018voodoo}
\APACinsertmetastar {%
whitson2018voodoo}%
\begin{APACrefauthors}%
Whitson, J\BPBI R.%
\end{APACrefauthors}%
\unskip\
\newblock
\APACrefYearMonthDay{2018}{}{}.
\newblock
{\BBOQ}\APACrefatitle {Voodoo software and boundary objects in game
  development: How developers collaborate and conflict with game engines and
  art tools} {Voodoo software and boundary objects in game development: How
  developers collaborate and conflict with game engines and art tools}.{\BBCQ}
\newblock
\APACjournalVolNumPages{new media \& society}{20}{7}{2315--2332}.
\PrintBackRefs{\CurrentBib}

\bibitem [\protect \citeauthoryear {%
Zhang%
\ \BBA {} Chen%
}{%
Zhang%
\ \BBA {} Chen%
}{%
{\protect \APACyear {2021}}%
}]{%
zhang2021using}
\APACinsertmetastar {%
zhang2021using}%
\begin{APACrefauthors}%
Zhang, Y.%
\BCBT {}\ \BBA {} Chen, J.%
\end{APACrefauthors}%
\unskip\
\newblock
\APACrefYearMonthDay{2021}{}{}.
\newblock
{\BBOQ}\APACrefatitle {Using Design Thinking in Educational Game Design: A Case
  Study of Pre-service Teacher Experience} {Using design thinking in
  educational game design: A case study of pre-service teacher
  experience}.{\BBCQ}
\newblock
\BIn{} \APACrefbtitle {International Conference on Blended Learning}
  {International conference on blended learning}\ (\BPGS\ 253--263).
\PrintBackRefs{\CurrentBib}

\bibitem [\protect \citeauthoryear {%
Zhu%
\ \BBA {} Wang%
}{%
Zhu%
\ \BBA {} Wang%
}{%
{\protect \APACyear {2019}}%
}]{%
10.1145/3365000}
\APACinsertmetastar {%
10.1145/3365000}%
\begin{APACrefauthors}%
Zhu, M.%
\BCBT {}\ \BBA {} Wang, A\BPBI I.%
\end{APACrefauthors}%
\unskip\
\newblock
\APACrefYearMonthDay{2019}{{\APACmonth{11}}}{}.
\newblock
{\BBOQ}\APACrefatitle {Model-Driven Game Development: A Literature Review}
  {Model-driven game development: A literature review}.{\BBCQ}
\newblock
\APACjournalVolNumPages{ACM Comput. Surv.}{52}{6}{}.
\newblock
\begin{APACrefURL} \url{https://doi.org/10.1145/3365000} \end{APACrefURL}
\newblock
\begin{APACrefDOI} \doi{10.1145/3365000} \end{APACrefDOI}
\PrintBackRefs{\CurrentBib}

\end{thebibliography}

\end{document}